\documentclass[a4paper]{article}

\usepackage{amsmath}
\usepackage{amssymb}
\usepackage{multirow}
\usepackage{tikz}
\usepackage{changepage}
\usepackage{pdflscape}
\usepackage{placeins}

\usepackage[top=7cm,bottom=2.5cm,headheight=120pt,headsep=15pt,left=6cm,right=1.5cm,marginparwidth=4.2cm,marginparsep=0.5cm]{geometry}

\usepackage{marginnote}
\reversemarginpar  
\usepackage{lipsum} 
\usepackage{calc}
\usepackage{siunitx}
\usepackage{lineno}
\usepackage[utf8]{inputenc}
\usepackage[T1]{fontenc}
\usepackage[
backend=biber,
natbib=true,
sortcites=true,
defernumbers=true,
style=authoryear,
citestyle=authoryear-comp,
maxnames=99,
maxcitenames=2,
giveninits=true,
terseinits=true,
url=false
]{biblatex}
\DeclareNameAlias{default}{family-given}
\renewbibmacro{in:}{%
  \ifentrytype{article}{}{\printtext{\bibstring{in}\intitlepunct}}} 
\DeclareFieldFormat[article]{pages}{#1} 
\AtEveryBibitem{\ifentrytype{article}{\clearfield{number}}{}} 
\DeclareFieldFormat[article, inbook, incollection, inproceedings, misc, thesis, unpublished]{title}{#1} 
\usepackage{csquotes}
\RequirePackage[english]{babel} 
\DeclareBibliographyCategory{ignore}
\addtocategory{ignore}{recommendation} 
\usepackage{nameref}
\usepackage[pdfborder={0 0 0}]{hyperref}  
\usepackage{graphbox}  
\usepackage{microtype}
\DisableLigatures[f]{encoding = *, family = * }
\RequirePackage[default,scale=0.90]{opensans}
\usepackage{xcolor}
\definecolor{darkgray}{HTML}{808080}
\definecolor{mediumgray}{HTML}{6D6E70}
\definecolor{ligthgray}{HTML}{d9d9d9}
\definecolor{pciblue}{HTML}{74adca}
\definecolor{opengreen}{HTML}{77933c}
\usepackage{changepage}
\usepackage[aboveskip=1pt,labelfont=bf,labelsep=period,singlelinecheck=off]{caption}

\usepackage{fancyhdr}  
\usepackage{lastpage}  
\fancyhfoffset[L]{4.5cm}  
\renewcommand{\headrulewidth}{\ifnum\thepage=1 0.5pt \else 0pt \fi} 

\newcommand{\DOIrecommendationlink}{\href{https://doi.org/\DOIrecommendation}{https://doi.org/\DOIrecommendation}}

\newcommand{\PCI}{Peer Community In Mathematical and Computational Biology}

\newcommand{\beginingpreprint}{
\vspace*{0.5cm}
\begin{flushleft}
\baselineskip=30pt
\marginpar{
\large\textnormal{\color{pciblue}\\RESEARCH ARTICLE}\\
\vspace*{0.5pt}
\\
\includegraphics[align=c,width=0.5cm]{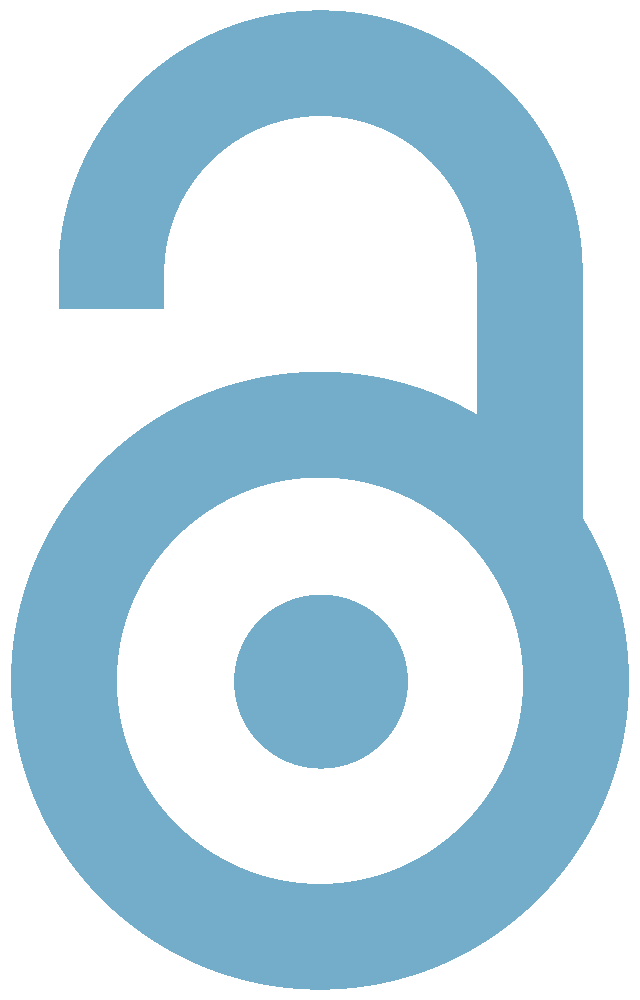} \space \large\textbf{\color{pciblue}Open Access}\\
\\
\includegraphics[align=c,width=0.5cm]{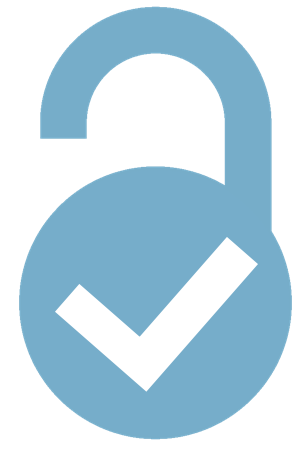} \space \large\textbf{\color{pciblue}Open Peer-Review}\\
\\
\includegraphics[align=c,width=0.5cm]{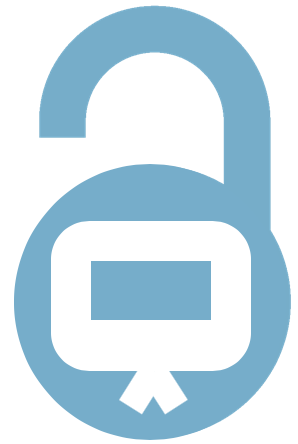} \space \large\textbf{\color{pciblue}Open Code}\\
\\
\\
\\
\\
\raggedright
\scriptsize\textbf{Cite as:}\space
\citeas\\
\vspace*{0.5cm}
\textbf{Posted:} \datepub\\
\vspace*{0.5cm}
\textbf{Recommender:}\\
\recommender\\
\vspace*{0.5cm}
\textbf{Reviewers:}\\
\reviewers\\
\vspace*{0.5cm}
\textbf{Correspondence:}\\
\href{mailto:\email}{\email}\\

}
{\Huge
\fontseries{sb}\selectfont{\preprinttitle}}
\end{flushleft}
\vspace*{0.25cm}
\begin{flushleft}

\Large
\listauthors
\end{flushleft}
\bigskip
{\raggedright
\listinstitutions}
\begin{flushleft}
\fcolorbox{lightgray}{lightgray}{
\parbox{\textwidth - 2\fboxsep}{
\centering\large{\fontseries{sb}\selectfont{This article has been peer-reviewed and recommended by\\
\emph{\PCI} (\DOIrecommendationlink)}}\\
}}
\end{flushleft}
\vspace*{0.5cm}
\fcolorbox{pciblue}{pciblue}{
\parbox{\textwidth - 2\fboxsep}{
\vspace{0.25cm}
\textbf{\large{\textsc{Abstract}}}\\
\preprintabstract\\

\footnotesize{\textbf{\emph{Keywords: }}\preprintkeywords}
\vspace{0.25cm}}
}
}

\newcommand{\preprinttitle}{The origin of the allometric scaling of lung ventilation in mammals}

\newcommand{\listauthors}{\raggedright 
Frédérique Noël\textsuperscript{1}, \space
Cyril Karamaoun\textsuperscript{1}, \space
Jerome A. Dempsey\textsuperscript{2} \&
Benjamin Mauroy\textsuperscript{3}
}

\newcommand{\listinstitutions}{
\textsuperscript{1} Université Côte d'Azur, LJAD, Vader center, Nice, France -- Nice, France 
\\
\textsuperscript{2} John Rankin Laboratory of Pulmonary Medicine, Department of Preventive Medicine, University of Wisconsin School of Medicine -- Madison, USA
\\
\textsuperscript{3} Université Côte d'Azur, CNRS, LJAD, Vader center, Nice, France -- Nice, France 
}

\newcommand{\datepub}{3 September 2021}

\newcommand{\recommender}{Wolfram  Liebermeister}

\newcommand{\DOIrecommendation}{10.24072/pci.mcb.100005}

\newcommand{\reviewers}{Oliver Ebenhöh, Stefan Schuster, Elad Noor and Megumi Inoue}

\newcommand{\citeas}{Noël F., Karamaoun C., Dempsey J. A. and Mauroy B. (2021) The origin of the allometric scaling of lung ventilation in mammals. arXiv, 2005.12362, ver. 6 peer-reviewed and recommended by Peer community in Mathematical and Computational Biology. https://arxiv.org/abs/2005.12362.}

\newcommand{\email}{benjamin.mauroy@univ-cotedazur.fr}

\newcommand{\preprintabstract}{A model of optimal control of ventilation has recently been developed for humans.
This model highlights the importance of the localization of the transition between a convective and a diffusive transport of respiratory gas.
This localization determines how ventilation should be controlled in order to minimize its energetic cost at any metabolic regime.
We generalized this model to any mammal, based on the core morphometric characteristics shared by all mammalian lungs and on their allometric scaling from the literature.
Since the main energetic costs of ventilation are related to convective transport, we prove that, for all mammals, the localization of the shift from a convective transport to a diffusive transport plays a critical role on keeping this cost low while fulfilling the lung function. 
Our model predicts for the first time the localization of this transition in order to minimize the energetic cost of ventilation, depending on mammal mass and metabolic regime. 
From this optimal localization, we are able to predict allometric scaling laws for both tidal volumes and breathing rates, at any metabolic rate. 
We ran our model for the three common metabolic rates -- basal, field and maximal -- and showed that our predictions reproduce accurately experimental data available in the literature. 
Our analysis supports the hypothesis that mammals allometric scaling laws of tidal volumes and breathing rates at a given metabolic rate are driven by a few core geometrical characteristics shared by mammalian lungs and by the physical processes of respiratory gas transport.}

\newcommand{\preprintkeywords}{Allometry; Respiratory system; Gas transport; Metabolic rate; Lung morphometry}

\newcommand{\vo}{\dot{V}_{O_2}}
\newcommand{\vom}{\dot{V}_{O_2}^{\text{max}}}
\newcommand{\vob}{\dot{V}_{O_2}^{\text{\rm{BMR}}}}
\newcommand{\vof}{\dot{V}_{O_2}^{\text{\rm{FMR}}}}
\renewcommand{\P}{\mathcal{P}}
\newcommand{\va}{\dot{V}_{A}}
\newcommand{\ve}{\dot{V}_E}
\newcommand{\Pb}{P_{\text{blood}}}
\newcommand{\dsp}{\displaystyle}

\begin{document}
\beginingpreprint


\section*{Introduction}

In animals, cellular respiration refers to the aerobic oxidation of fatty acids and glucose that represents a major source of energy production \parencite{lodish_molecular_2008}. 
Oxidative processes require oxygen to be brought from the atmosphere to each individual cell.
In parallel, carbon dioxide, a major by-product of cellular respiration, has to be removed from the tissues \parencite{terjung_lung_2016}.
Capture and transport of oxygen and removal of carbon dioxide is performed by the respiratory and circulatory systems.
The lung handles the transport of oxygen from the ambient air to the alveolar exchange surface, which is in contact with the blood network. 
Then, the circulatory system transports oxygen from the lung exchange surface to cells.
Conversely, carbon dioxide is transported from cells to the ambient air \parencite{west_respiratory_2011}.

The mammalian lung has been selected and shaped by evolution to fulfill the body needs in oxygen and to eliminate carbon dioxide \parencite{terjung_lung_2016}.
It is composed of two main parts: the bronchial tree and the respiratory zone. 

The bronchial tree is structured as a nearly dichotomous tree, where an airflow circulates during the process of ventilation, which consists in a succession of inspiration and expiration cycles. 
At inspiration, fresh air is brought into the respiratory zone, where oxygen exchange with blood takes place. 
In parallel, carbon dioxide is transferred from blood to alveoli.
Then, at expiration, a higher carbon dioxide/lower oxygen air is expelled from the lung~\parencite{west_respiratory_2011}.

The respiratory zone forms a large and thin exchange surface between alveolar air and blood.
This surface is folded into the thorax cavity and connected to ambient air thanks to a compact bronchial tree.
These characteristics have evolved to fulfill the gas exchange requirements in mammals while satisfying the structural body needs, i.e. a compact and rib-covered thorax cavity~\parencite{mauroy_optimal_2004}.

The transport of air in the lung by ventilation requires energy.
A hydrodynamic resistance to the air flow in the bronchi arises from friction effects, due to air viscosity~\parencite{mauroy_viscosity_2014}. 
In parallel, mechanical energy is needed to expand the thoracic cage and the lung tissues during inspiration. 
That energy is lost at expiration by the viscoelastic recoil of the tissues, at least at rest \parencite{west_respiratory_2011}. 
Without a careful regulation, these physical constraints could have a high metabolic cost, even at rest~\parencite{otis_mechanics_1950}. 
However, natural selection favors configurations that require low amounts or minima of energy. 
Moreover, the process of optimization by evolution is performed under the constraint of the lung function: the gas exchanges have to fit the metabolic activity requirements.

The typical functional constraint associated to this energy cost was up to recently based on the total air flow rate entering the lung only~\parencite{otis_mechanics_1950, mead_control_1960, johnson_biomechanics_2007}, without accounting for the respiratory gas transport and the gas exchange requirements. More recently, \textcite{noel_interplay_2019} optimized the energy spent for ventilation in humans with a more realistic functional constraint, based on the oxygen flow in the alveoli, including the physics of oxygen and carbon dioxide transport in a symmetric branched model of lung. This approach was not only able to predict physiological ventilation parameters for a wide range of metabolic regimes, but it also highlighted the distribution and transport of oxygen and carbon dioxide in the lung.

Actually, the progression of air in the lung is a combination of two mass transport processes: convection and diffusion. 
In the upper and central part of the bronchial tree, the convective transport largely dominates the mass transport, driven by the pressure gradient imposed by the airflow. 
However, as the cumulative surface of the bronchi section area increases at each bifurcation, the air velocity decreases while progressing towards the deeper part of the tree. 
At some point, the characteristic velocity of convection becomes smaller than the characteristic velocity of diffusion;  
the mass transport becomes dominated by the diffusion process.
The localization of the transition zone between convection and diffusion depends on the geometry of the lung and on the ventilation parameters. 
The previous work of \textcite{noel_interplay_2019} showed that the control of ventilation in humans localizes the transition zone based on a trade-off between the oxygen demand and the availability and accessibility of the exchange surface deeper in the lung~\parencite{sapoval_smaller_2002, noel_interplay_2019}.

The lungs of mammals share morphological and functional properties, raising the question on whether the previous results for human can or cannot be extended to all mammals. 
These properties are known to depend on the mass $M$ of the animal, expressed in kg in this study, with non trivial power laws called allometric scaling laws~\parencite{huxley_terminology_1936,gunther_dimensional_1975,peters_ecological_1986,west_general_1997,terjung_lung_2016}.
The physics of ventilation, and hence its control, is linked to the lung geometry. Consequently, the morphological differences amongst mammals also affect the control of ventilation. 
This is supported by the allometric scaling laws followed by the ventilation frequency and tidal volume. 
Breathing rate at basal metabolic rate (BMR) has been estimated to follow the law $f_{b}^{\text{\rm{BMR}}} \simeq 0.58 \ M^{-\frac14} \ \rm{Hz}$~\parencite{worthington_relationship_1991} and tidal volume to follow the law $V_{T}^{\text{\rm{BMR}}} \simeq 7.14 \ M^1 \ \rm{mL}$~\parencite{west_general_1997,haverkamp_physiologic_2005}. 
At other metabolic rates, less data is available in the literature except for the breathing rate of mammals at maximal metabolic rate (MMR), estimated to follow the law $f_{b}^{\text{\rm{MMR}}} \simeq 5.08 \ M^{-0.14} \ \rm{Hz}$~\parencite{altringham_power_1991}. 
The links between these allometric scaling laws and the optimization of the energy spent for ventilation by mammals remains still to be uncovered. 
A model able to predict these laws for mammals would be a powerful tool to derive them at other regimes, such as at submaximal exercise, at maximal exercise or at field metabolic rate (FMR).

Actually, the ventilation properties at intermediate metabolic rates are difficult to obtain, making the study of the metabolism of mammals at these regimes difficult to analyse~\parencite{speakman_history_1998}. 
Hence, a clear biophysical understanding of the origin of these scaling laws could allow to extend ventilation-related analyses performed for one mammal species to another.
This could improve the pertinence of using animal models~\parencite{matute-bello_animal_2008, rocco_what_2020} or, to the contrary, of using human data, richer in the literature, to study the metabolism of other mammals ~\parencite{haverkamp_physiologic_2005}.

In this work, we develop two mathematical models: one to estimate the amount of oxygen captured from air by mammalian lungs; and one to estimate the energetic cost of ventilation. 
These two models depend on mammals mass and are coupled together to form a mathematical model for the natural selection of breathing rates and tidal volumes. 
In the frame of our model hypotheses, we show that the physiological allometric scaling laws reported in the literature for both breathing rates and tidal volumes are actually minimizing the mechanical energy of breathing. 
Moreover, we show that the selected configurations at a given metabolic rate are mainly driven by the geometries of the mammalian lungs and by the physical processes involved in oxygen transport in the lung.

\section*{Modelling}

The methodology and hypotheses used to perform our analysis are summarised in Tables \ref{generalApp}, \ref{model1Hyp} and \ref{model2Hyp} in Appendix \ref{VII}.
The derivation of the allometric properties of ventilation is based on the previous model developed by \textcite{noel_interplay_2019}, which is adapted to all mammals over 5 orders of magnitude in mass.

\subsection*{Ventilation pattern and energy cost of ventilation}

The cost of ventilation is estimated as in \textcite{noel_interplay_2019} and its computation is based on \textcite{otis_mechanics_1950, mead_control_1960, johnson_biomechanics_2007}.
The estimation of the cost is generalized to all mammals using the allometric scaling laws of the mechanical parameters involved in lung ventilation.

The velocity $u$ of the air entering the lung is represented by a sinusoidal pattern in time, i.e. 
\begin{equation}
u(t) = U \sin(2 \pi t / T)
\label{velocity}
\end{equation}
The quantity $U$ is the maximal velocity and $T$ is the period of ventilation and the inverse of the breathing frequency $f_b = 1/T$. 
Denoting $S_0$ the surface area of the tracheal cross-section, the tidal volume is then $V_T = U S_0 T / \pi$, see 
Appendix \ref{fn1} and the air flow rate is $\ve = V_T f_b$. 

The biomechanics of the lung ventilation involves two active physical phenomena that are the sources of an energy cost~\parencite{johnson_biomechanics_2007,noel_interplay_2019}. 
First, the motion of the tissues out of their equilibrium implies that the diaphragm has to use, during inspiration, an amount of energy that is stored in the tissues as elastic energy. 
This energy is then used during expiration for a passive recoil of the tissues. 
The power spent is related to the elastic properties of the thoracic cage and of the lung. 
These properties depend on the lung compliance $C$~\parencite{agostini_postural_2011}, which is defined as the ratio between the change in lung volume and the change in pleural pressure.
We derive the resulting power in Appendix \ref{fn2}.
Second, the airflow inside the bronchi induces an energy loss due to viscous effects that have to be compensated by the motion of the diaphragm during inspiration. 
The dissipated viscous power depends on the hydrodynamic resistance $R$ of the lung.
Details about the derivation of the resulting power are given in Appendix \ref{fn3}.

The total power $\tilde{\mathcal{P}}(V_T,f_b)$ spent by ventilation is the sum of these two powers 
\begin{equation}
\tilde{\mathcal{P}}(V_T,f_b) = \underbrace{\frac{V_T^2 f_b}{2 C}}_{\text{elastic power}} + \underbrace{\frac14\left(\pi f_b V_T \right)^2 R}_{\text{viscous power}}
\label{power}
\end{equation}
The compliance and the hydrodynamic resistance of the lung follows allometric scaling laws that have been derived at BMR: $C \propto M^1$~\parencite{stahl_scaling_1967} and $R \propto M^{-\frac34}$~\parencite{west_general_1997}. 
Nevertheless, lung volumes at exercise tend to stay within the linear part of the pressure--volume curve, suggesting that the compliance does not change much at exercise~\parencite{henke_regulation_1988}.
Also, the diameters of the airways adjust during exercise and maintain the lung resistance close to its rest value~\parencite{johnson_mechanical_1992}, see details in Appendix \ref{VI}. 
Hence, in our model, both the compliance and the hydrodynamic resistance are assumed independent of the metabolic regime. 
However, these hypotheses might not hold at very high exercise, where the power spent for ventilation is drastically increased due to non-linear responses. 
Hence, the previous hypotheses might underestimate the mechanical power needed for ventilation at intense exercise~\parencite{agostoni_static_2011, mauroy_interplay_2003}. 

The mechanical power has to be minimized with a constraint on the oxygen flow to blood $f_{O_2}(V_T, f_b)$, which has to match the oxygen flow demand $\vo$.
For a given mammal mass, the mathematical formulation of this optimization problem is 
\begin{equation}
\begin{aligned}
\mathop{\text{Min}}_{(V_T,f_b) \in \mathcal{H}}&\tilde{\mathcal{P}}(V_T, f_b)\\
& \text{ with } \mathcal{H} = \left\{ (V_T,f_b) \, | \, f_{O_2}(V_T, f_b) = \vo \right\}
\end{aligned}
\end{equation}
In the following, we estimate the oxygen flow $f_{O_2}(V_T, f_b)$ transferred to the blood during the ventilation and the oxygen flow demand $\vo$ according to the metabolic regime. 
Due to the complexity of the model, the optimization of the mechanical power is carried out numerically.
Our model predicts that the optimal tidal volume $V_T$ and breathing rate $f_b$ follow an allometric scaling law. 
\begin{landscape}
\begin{table*}[p]
\scriptsize
\centering
\begin{tabular}{cllll}
\hline
&\multirow{2}*{Variables} & \multicolumn{2}{c}{Exponent} & \multirow{2}*{Prefactor} \\
\cline{3-4}
 && Predicted \parencite{west_general_1997} & Observed & \\
\hline
\multirow{7}*{Morphometry}& 
$V_L$: Lung volume  & 1 & 1.06 \parencite{stahl_scaling_1967} & 53.5 mL \parencite{stahl_scaling_1967}\\
 &$r_0$: Tracheal radius & 3/8 (= 0.375)  & 0.39 \parencite{tenney_comparative_1967} & 1.83 $\text{mm}^*$ \\
 & $l_0$: Tracheal length & 1/4 (= 0.25) & 0.27 \parencite{tenney_comparative_1967} & 1.87 $\text{cm}^*$\\
 &$r_A$: Radius of alveolar ducts & 1/12 ($\simeq$ 0.083) & 0.13 \parencite{tenney_quantitative_1970} & 0.16 $\text{mm}^*$ \\
   &$l_A$: Length of alveolar ducts & -1/24 ($\simeq$ -0.042) & N.D. & 1.6 $\text{mm}^*$\\
& $n_A$: Number of alveoli  & 3/4 (= 0.75) & N.D. & 12 400 000$^*$ \\
 &$v_A$: Volume of alveolus  & 1/4 (= 0.25) & N.D. & N.D.\\ \hline
 \multirow{6}*{Physics}& $f_b$ : Respiratory frequency (rest)  & -1/4 (= -0.25) & -0.26 \parencite{stahl_scaling_1967} & 53.5 $\text{min}^{-1}$ \parencite{stahl_scaling_1967}\\
&$V_T$: Tidal volume (rest) & 1 & 1.041 \parencite{west_general_1997} & 7.69 mL \parencite{stahl_scaling_1967} \\
&$P_{50}$: $O_2$ affinity of blood  & -1/12 ($\simeq$ -0.083) & -0.089 \parencite{dhindsa_comparative_1971} & 37.05 $\text{mmHg}^*$ \\
&$R$: Total resistance  & -3/4 (= -0.75) & -0.70 \parencite{stahl_scaling_1967} & 24.4 $\text{cmH$_2$O} \,\, \text{s} \, \text{L}^{-1} $ \parencite{stahl_scaling_1967}\\
&$C$: Total compliance  & 1 & 1.04 \parencite{stahl_scaling_1967} & 1.56 mL $\text{cmH$_2$O}^{-1}$ \parencite{stahl_scaling_1967} \\ 
&$P_{\rm{pl}}$: Interpleural pressure  & 0 & 0.004 \parencite{gunther_physiometry_1966} & N.D.\\
\hline
\multicolumn{1}{c}{}                         & \multicolumn{1}{c}{}       &  \multicolumn{1}{c}{} & \multicolumn{1}{c}{}\\ 
& \multicolumn{1}{l}{Variables}   & \multicolumn{1}{l}{Exponent at BMR} & \multicolumn{1}{l}{Exponent at FMR} & \multicolumn{1}{l}{Exponent at MMR}\\  \hline
\multirow{2}*{Metabolism} & $ \dsp  \vo$: $O_2$ consumption rate & 3/4 (= 0.75) \parencite{peters_ecological_1986, kleiber_body_1932} & 0.64 \parencite{hudson_relationship_2013}& 7/8 (= 0.875) \parencite{weibel_exercise-induced_2005} \\
& \multicolumn{1}{p{3cm}}{$t_c$: Transit time of blood in pulmonary capillaries}& \multicolumn{1}{p{2.5cm}}{1/4 (= 0.25) \parencite{west_general_1997,haverkamp_physiologic_2005}}& 1/4 (= 0.25) (hypothesized) &0.165 \parencite{haverkamp_physiologic_2005,bishop_integration_2013} \\
\hline
\end{tabular}
\caption{Predicted and observed/computed values of allometric exponents for the mammalian respiratory system. $^*$: Prefactor computed using human values (M = 70 kg) at rest and computed for masses expressed in kg.  BMR: Basal Metabolic Rate, FMR: Field Metabolic Rate, MMR: Maximal Metabolic Rate. N.D.: No data found.}
\label{tab:allo_expo}
\end{table*}
\end{landscape}

\subsection*{Core characteristics of the geometry of the mammalian lung}
 
\begin{figure}[h!]
\begin{center}
\includegraphics[width=8cm]{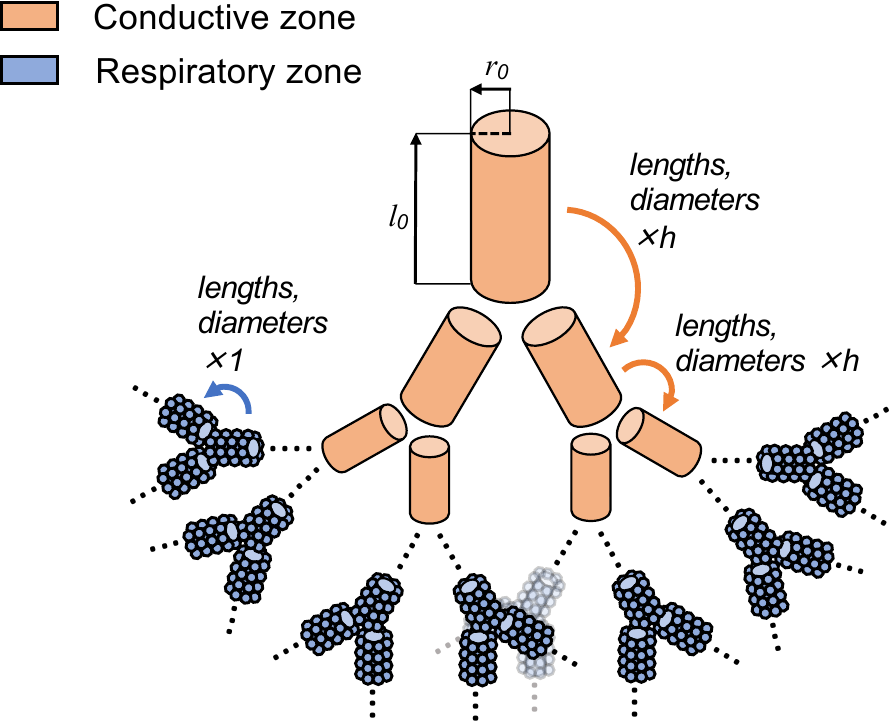}
\caption{Illustration of the lung model used in this work.
Our model is based on the assembly of self-similar trees with symmetric bifurcations that mimic the two functional zones. 
The tree in beige mimics the bronchial tree, where oxygen is only transported along the branches. 
The tree in blue mimics the acini, where oxygen is transported along the branches and also captured in the alveoli that cover the walls of the branches.
The model of oxygen transport depends only on airways lengths and diameters but not on airways spatial distribution in the lungs.
Hence, branching angles and branching planes orientations are not accounted for in this work.}
\label{fig:tree}
\end{center}
\end{figure}

The lungs of mammals share invariant characteristics~\parencite{weibel_pathway_1984}. 
First, the lung has a tree-like structure with bifurcating branches. 
It decomposes into two parts: the bronchial tree or conductive zone that transports, mainly by convection, the (de)oxygenated air up and down the lung, and the acini or respiratory zone, where gas exchanges with blood occur through the alveolar--capillary membrane. 
The bronchial tree can be considered as self-similar, as the size of its branches is decreasing at each bifurcation with a ratio close to $h = \left(\frac12\right)^{\frac13}$~\parencite{weibel_pathway_1984, mauroy_optimal_2004, karamaoun_new_2018}. 
In the acini, the size of the branches are almost invariant at bifurcations~\parencite{weibel_pathway_1984, tawhai_ct-based_2004}.
Thus, the bronchial tree and the acini are modelled as airway trees with symmetric bifurcations~\parencite{mauroy_optimal_2004, mauroy_toward_2011, mauroy_toward_2015, noel_interplay_2019}, as shown in Figure \ref{fig:tree}.
This model accounts for the lung branching pattern and for the lengths and diameters of airways, but not for the airways spatial distribution.
Actually, the properties such as branching angles and orientations of the branching planes are not relevant in the model of oxygen transport developed in this work.
Moreover, some mammals species have specific branching pattern~\parencite{raabe_tracheobronchial_1976, maina_morphometric_2001, metzger_branching_2008}. 
However, we only retain in our model the core property of the mammalian lungs: the tree-like structure. 

A generation of the tree corresponds to the set of branches for which the path to the root of the tree, which mimics the trachea, contains the same number of bifurcations. 
The bronchial tree is modelled with $G$ successive generations. 
The branches in the generation $G$ are connected to the acini.
The acini are modelled with $H$ generations~\parencite{weibel_pathway_1984}. 
The total number of generations of the tree is then $N = G+H$. 
Hence, if the radius and length of the root of the tree are $r_0$ and $l_0$, the radius $r_i$ and length $l_i$ of an airway in the generation $i$ is
\begin{equation}
\left.
\begin{array}{llll}
r_i = 
\left\{
\begin{array}{ll}
r_0 h^i & i=0...G-1\\
r_{G-1} & i=G...N-1\
\end{array}
\right.
\hspace{0.5cm}
l_i = 
\left\{
\begin{array}{ll}
l_0 h^i & i=0...G-1\\
l_{G-1} & i=G...N-1
\end{array}
\right.
\end{array}
\right.
\label{selfsim}
\end{equation} %
In the generation $i$, the airways surface area $S_i$ and the mean air velocities $u_i$ in the airways are related to the generation index $i$ and can be computed from the airways radii scaling laws, see Appendix \ref{fn4}.

The derivation of a lung model that depends only on mammal mass requires to relate explicitly the morphological parameters involved in our model with the animal mass. 
We used the datasets from \textcite{west_general_1997}.
These authors derived for the cardiorespiratory system many theoretical allometric scaling laws that are in good agreement with ecological observations.
 
The morphological parameters used in our model are the trachea radius $r_0$, the reduced trachea length $l_0$, the generations number $G$ of the bronchial tree, the generations number $H$ of the acini and the amount $\rho_s$ of exchange surface area per unit of surface area of the alveolar duct wall.
Each of these quantities follows an allometric scaling law that can be derived from \textcite{west_general_1997}: 
\begin{itemize}
\item The radius $r_0$ of the trachea scales as $M^{\frac38}$~\parencite{west_general_1997}.
The bronchi radii, and consequently the dead volume, are affected by the ventilation regime~\parencite{johnson_mechanical_1992, dempsey_respiratory_2015}. 
The airways radii in our lung model are computed from the tree root radius $r_0$, see equation (\ref{selfsim}).
Hence, the dependence on metabolic rate of the dead volume is integrated into the prefactor of the tracheal radius allometric scaling law, see Appendix \ref{fn5}.
\item The allometric scaling law for the tracheal length $l_0 \propto M^{\frac14}$ can be derived from \textcite{west_general_1997}, see Appendix \ref{fn6}. 
\item Based on the hypothesis that the radii of alveolar ducts are similar to the radii of aveoli $r_A$~\parencite{weibel_pathway_1984} and based on the allometric scaling law $r_A \propto M^{\frac1{12}}$~\parencite{west_general_1997}, we can deduce that $2^G \propto M^{\frac78}$ and hence determine $G$, see Appendix \ref{fn7}.
Moreover, from \textcite{rodriguez_pulmonary_1987, haefeli-bleuer_morphometry_1988}, we can assume that the number of generations of alveolar ducts $H$ in the acini is independent of the mammal mass and equal to $6$.
\item Then, relating the scalings of $G$ and $H$ to the allometric scaling law for the exchange surface area $S_A \propto M^{\frac{11}{12}}$~\parencite{west_general_1997}, we deduce that the amount of exchange surface area per unit of alveolar duct surface area $\rho_s$ is independent of the mammal mass, see Appendix \ref{fn8}.
\end{itemize}

 Hence, the number of airway generations $N=G+H$ predicted is about $13$ for a $30$ g mouse and about $23$ for a $70$ kg human, in agreement with physiological data~\parencite{gomes_geometric_2002, weibel_morphometry_1963}. 

\subsection*{Oxygen transport and exchange with blood}

The oxygen transport and exchange model in the human lung from \textcite{noel_interplay_2019} is extended to any mammal, based on its mass $M$.
The transport and exchange now occur in the idealized lung that has been generalized in the previous section to fit any mammal.
The parameters of the transport and exchange model from \textcite{noel_interplay_2019} are also adjusted using relevant allometric scaling laws from \textcite{west_general_1997}.

The transport of oxygen in the lung is driven by three phenomena: convection by the airflow, diffusion and exchange with blood through the alveoli walls. 
The partial pressure of oxygen averaged over the section of an airway is transported along the longitudinal axis $x$ of the airway.
In the alveolar ducts, the oxygen exchange with blood occurs in parallel with the oxygen transport. 
Hence, in each airway belonging to the generation $i$, the partial pressure of oxygen follows the convection--diffusion--reaction equation derived in \textcite{noel_interplay_2019} and in Appendix \ref{I},
\begin{multline}
\frac{\partial P_{i}}{\partial t} - \underbrace{D \frac{\partial ^2 P_{i}}{\partial x^2}}_{\text{diffusion}} + \underbrace{u_{i}(t)\frac{\partial P_{i}}{\partial x}}_{\text{convection}} + \underbrace{\beta_i \left(P_{i} - P_{\text{blood}} \right)}_{\text{exchange with blood}} = 0, \\
\text{ for } x \in [0,l_i],
\label{eq:eq_1}
\end{multline} 
where $P_i$ is the mean oxygen partial pressure over the airway section, $D$ is the oxygen diffusion coefficient in air and $u_i(t)$ is the mean air velocity in an airway of generation $i$. 
The reactive term $\beta_i$ mimics the exchanges with blood through the airway wall.
The quantity $\beta_i$ is equal to zero in the convective tree ($i=0...G-1$) and is positive in the acini ($i=G...N-1$). 
In the acini, the oxygen exchange occurs through the wall of the ducts and $\beta_i$ depends on the membrane and oxygen chemical properties, on the membrane thickness and on the local exchange surface derived from $\rho_s$, see Appendix \ref{fn8}. 
More details about the derivation of $\beta_i$ are given in Appendix \ref{I}.
As a consequence, the reaction term $\beta_i$ follows an allometric scaling law, $\beta_i \propto M^{\frac1{12}}$, see Appendix \ref{fn9}.

To determine the oxygen partial pressure in blood that drives the oxygen exchange, we assume that the flow of oxygen leaving an alveolar duct through its corresponding exchange surface is equal to the flow of oxygen that is captured by blood, accounting for the oxygen dissolved in the blood plasma and for the oxygen captured by haemoglobin~\parencite{felici_physics_2003, noel_interplay_2019}, see Appendix \ref{II} for more details. 

The bifurcations are mimicked using boundary conditions that connect a generation to the next: we assume that the partial pressures are continuous at the bifurcations and that the amount of oxygen that goes through the bifurcation is conserved, see Appendix \ref{III}. 

Finally, the system is initialised at the time $t=0$ s using a distribution of partial pressures detailed in Appendix \ref{IV}. 

With these hypotheses, our model takes as inputs the mass of the mammal $M$, the oxygen flow needed by the body $\vo$, the tidal volume $V_T$ and the breathing frequency $f_b$.
The model outputs the flow of oxygen $f_{O_2}(V_T, f_b)$ exchanged with blood, see Appendix \ref{fn10}.

\subsection*{Power optimization with a constrained oxygen flow}

We search for the minimum of $\tilde{\mathcal{P}}(V_T,f_b)$ relatively to the tidal volume $V_T$ and the breathing frequency $f_b$, see equation (\ref{power}).
The minimization is made with a constraint on the oxygen flow to blood, written mathematically $f_{O_2}(V_T,f_b) = \vo$. 
The quantity $f_{O_2}(V_T,f_b)$ is the oxygen flow to blood resulting from a lung ventilation with the characteristics $(V_T, f_b)$ and estimated with our model of oxygen transport and exchange in the mammalian lung.
The quantity $\vo$ is the oxygen flow needed by the metabolism at the regime considered. 
Allometric scaling laws for mammals of basal, field and maximal metabolic rates are available in the literature, see Table \ref{metabo}.
With these scalings, we can compute the desired oxygen flow $\vo$ depending on the animal mass $M$ and on the metabolic regime.
\begin{table}[t]
\centering
\begin{tabular}{lll}
\hline
 Metabolic rate & Allometric scaling law & Reference\\
 \hline
Basal (BMR) & $\vob \propto M^{\frac34}$ & \parencite{kleiber_body_1932, peters_ecological_1986}\\
Field (FMR) & $\vof \propto M^{0.64}$ & \parencite{hudson_relationship_2013}\\
Maximal (MMR) & $\vom \propto M^{\frac78}$ & \parencite{weibel_exercise-induced_2005}\\
\hline
\end{tabular}
\caption{Allometric scaling laws for mammals of the needed oxygen flow $\vo$ at three metabolic rates. $M$ is the mammal mass. Data from the literature.}
\label{metabo}
\end{table} 
Other exponents for metabolic rates, less pertinent for our study, have also been derived for specific subsets of mammals species, based for example on their size or on their athletic capacity~\parencite{white_mammalian_2003, weibel_allometric_2004}.

The resolution of the model equations and the optimization process are performed using numerical simulations, as in \textcite{noel_interplay_2019}. 
The numerical strategy is described in Appendix \ref{V} and details about the sensitivity of the model to its parameters are given in Appendix \ref{VI}.
The software and its details are available in the open data repository Zenodo~\parencite{noel_code_2021}.

%


\section*{Results}

Our analysis assumes that mammals evolution selected for the minimum of the mechanical cost of ventilation while allowing the lung to fulfill its functions of oxygen transfer to blood. 

Our modelling approach mimics this process and allows to determine optimal values for the breathing rate $f_b$ and the tidal volume $V_T$ from the mass $M$ of a mammal and from its metabolic rate.
The mechanical power of ventilation $\tilde{\P}(V_T,f_b)$, estimated in equation (\ref{power}), is optimized with a constraint on the oxygen flow.
This functional constraint is expressed in our model as $f_{O_2}(V_T,f_b) = \vo$.
The oxygen flow $f_{O_2}(V_T,f_b)$ is computed using our model of oxygen transport and exchange in an idealised lung, see Figure \ref{fig:tree} and equations (\ref{eq:eq_1}).  
The quantity $\vo$ is the targeted oxygen flow and corresponds to the mean oxygen demand for a mammal of mass $M$ at the metabolic regime studied. 
Basal, field and maximal metabolic rates are analysed and the corresponding $\vo$ allometric scalings are determined from the literature, see Table \ref{metabo}.

A synthesis of the hypotheses of our models is given in Appendix \ref{VII} in Tables \ref{generalApp}, \ref{model1Hyp} and \ref{model2Hyp}.

\begin{figure*}[htp!]
\begin{center}
\phantom{}\newline
{\Large A} \includegraphics[width=10cm]{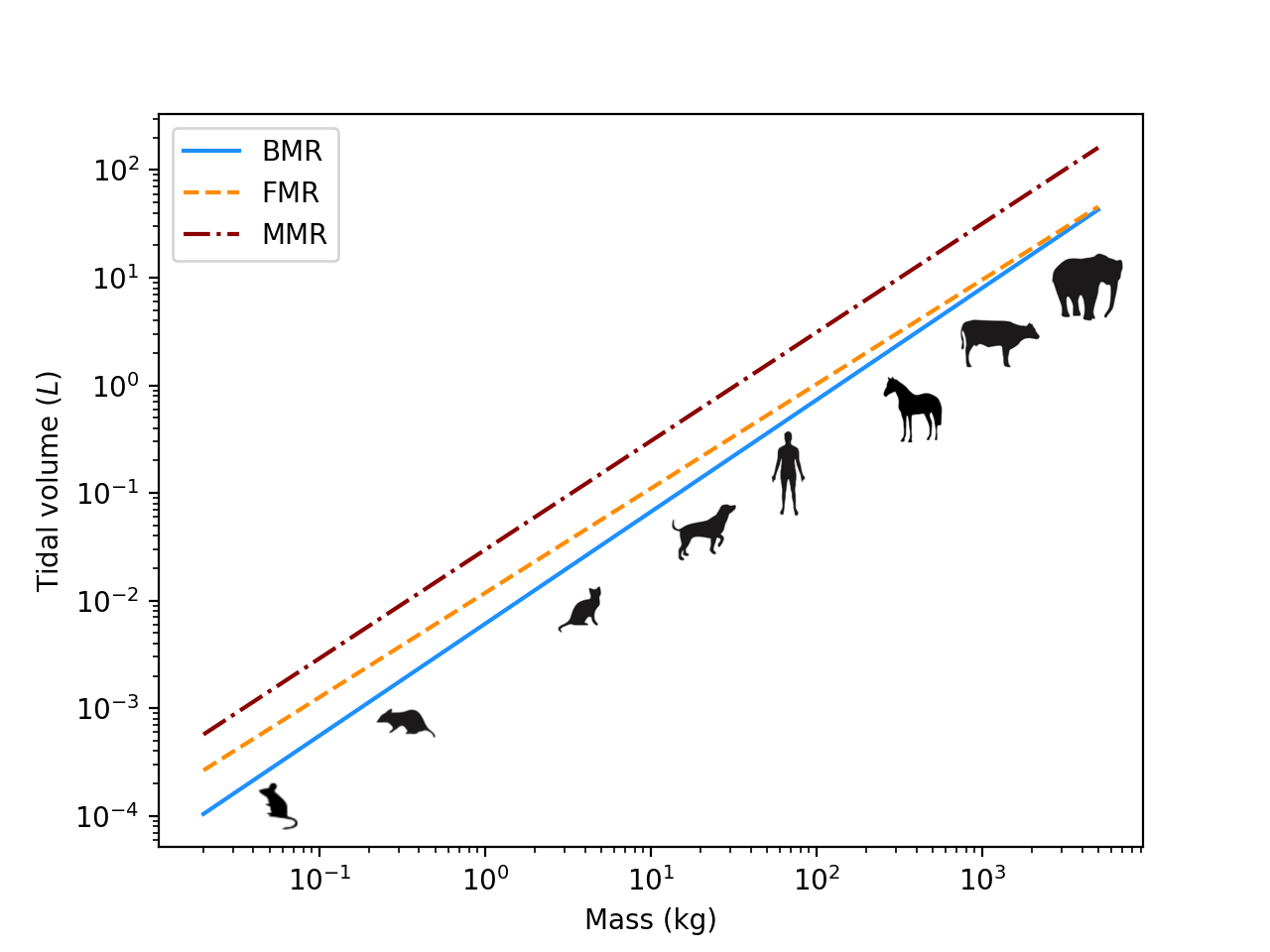}
\newline
{\Large B} \includegraphics[width=10cm]{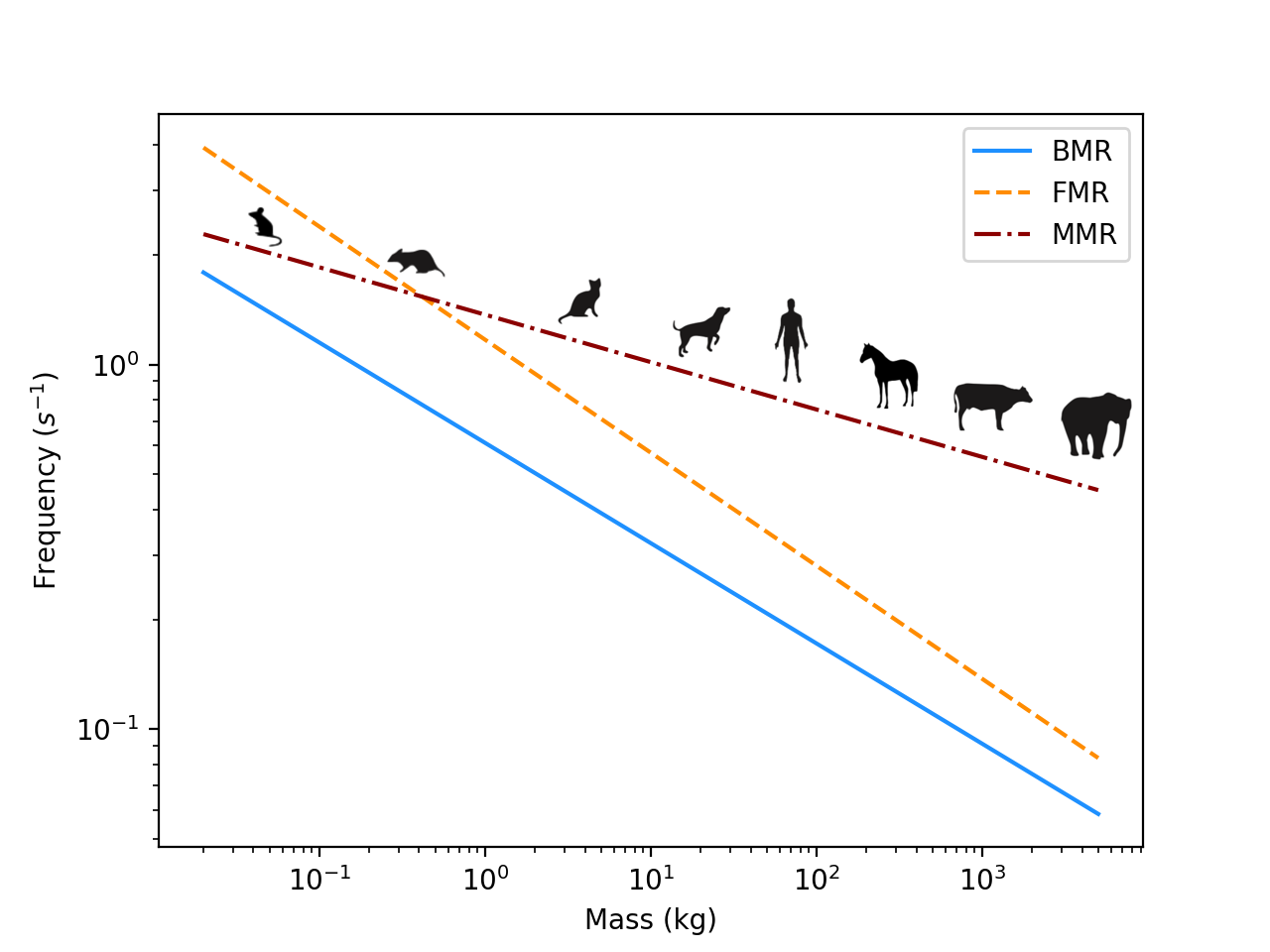}
\newline
\caption{{\bf A:} Predicted tidal volume as a function of the mammal mass (log-log). 
Solid line: BMR, $V_{T}^{\text{\rm{BMR}}} \simeq 6.1 \ M^{1.04} \ \text{ml}$; dashed line: FMR, $V_{T}^{\text{\rm{FMR}}} \simeq 11.8 \ M^{0.97} \ \text{ml}$, dash-dotted line: MMR, $V_{T}^{\text{\rm{MMR}}} \simeq 29.7 \ M^{1.01} \ \text{ml}$. 
{\bf B:} Predicted breathing frequency as a function of the mammal mass (log-log). 
Solid line: BMR, $f_{b}^{\text{\rm{BMR}}} \simeq 0.61 \ M^{-0.27} \ \text{Hz}$; dashed line: FMR, $f_{b}^{\text{\rm{FMR}}} \simeq 1.17 \ M^{-0.31} \ \text{Hz}$, dash-dotted line: MMR, $f_{b}^{\text{\rm{MMR}}} \simeq 1.37 \ M^{-0.17} \ \text{Hz}$. 
A larger dead volume at exercise~\parencite{haverkamp_physiologic_2005} makes the oxygen source for diffusion slower to deplete.
This might lead to a decrease in the optimal breathing rate, depending on the increase of the oxygen need. 
As a consequence, for small mammals, our model predicts breathing frequencies at MMR smaller than breathing frequencies at FMR. 
}
\label{fig:volM}
\label{fig:freqM}
\end{center}
\end{figure*}

\subsection*{Allometric scaling laws of breathing rates and tidal volumes}

In 1950, Otis et al. optimized $\tilde{\P}(V_T,f_b)$ with a constraint on the alveolar ventilation $\va = (V_T - V_D) f_b$, where $V_D$ is the dead volume.
They showed that an optimal breathing frequency could be computed analytically~\parencite{otis_mechanics_1950, johnson_biomechanics_2007}.
Using data available in the literature~\parencite{gunther_dimensional_1975, stahl_scaling_1967, west_general_1997, haverkamp_physiologic_2005} and the analytic formula from Otis et al., we derived allometric scaling laws for breathing frequency and tidal volume at BMR, $f_{b,\rm{pred}}^{\rm{\rm{BMR}}} = 0.9 \ M^{-\frac14} \ \rm{Hz}$ and $V_{T,\rm{pred}}^{\rm{\rm{BMR}}} =  7.5 \ M^1 \ \rm{ml}$, see Appendix \ref{fn11}.
The computed allometric scaling laws are in good agreement with observations, supporting the minimal ventilation mechanical power hypothesis.
However, this approach is not able to predict allometric laws at regimes other than BMR.
Actually, the localization of the convection--diffusion transition in the lung drives the amount of oxygen flow to blood~\parencite{noel_interplay_2019}.
Hence, only a model that is able to localize this transition in the tree and to compute precisely the amount of oxygen exchange would be able to reach satisfactory predictions. 

We ran our model for the three metabolic regimes BMR, FMR and MMR.
It predicts that breathing rates and tidal volumes follow allometric scaling laws in all the three regimes, see Figure \ref{fig:volM},
\begin{equation}
\begin{array}{ll}
\hspace{0.5cm}f_{b}^{\text{\rm{BMR}}} \simeq 0.61 \ M^{-0.27} \ \text{Hz} \text{,} & V_{T}^{\text{\rm{BMR}}} \simeq 6.1 \ M^{1.04} \ \text{ml}\\

\hspace{0.5cm}f_{b}^{\text{\rm{FMR}}} \simeq 1.17 \ M^{-0.31} \ \text{Hz} \text{,} & V_{T}^{\text{\rm{FMR}}} \simeq 11.8 \ M^{0.97} \ \text{ml}\\

\hspace{0.5cm}f_{b}^{\text{\rm{MMR}}} \simeq 1.37 \ M^{-0.17} \ \text{Hz} \text{,} & V_{T}^{\text{\rm{MMR}}} \simeq 29.7 \ M^{1.01} \ \text{ml}
\end{array}
\label{results}
\end{equation}
Our model predicts exponents that are in accordance with the values observed in the literature, see Table \ref{allom}.
Moreover, the predicted prefactors show that our model is able to give quantitative predictions in accordance with the physiology of the mammalian lungs.  

\begin{table}[t]
\centering 
\begin{tabular}{lllll}
\hline
& $f_b$ (pred.) & $f_b$ (obs.) & $V_T$ (pred.) & $V_T$ (obs.)\\
\hline
BMR &
 $-0.27$ & $-0.25$ & $1.04$ & $1$\\
FMR&
$-0.31$ & N.D. & $0.97$ & N.D.\\
MMR &
$-0.17$ & $-0.14$ & $1.01$ & N.D.\\
\hline
\end{tabular} 
\caption{Predicted and observed exponents for the allometric scaling laws of breathing frequency $f_b$ and tidal volume $V_T$ at three different metabolic regimes~\parencite{worthington_relationship_1991, altringham_power_1991, west_general_1997, haverkamp_physiologic_2005}, see equations (\ref{results}).}
\label{allom}
\end{table}

\subsection*{Transition between convection and diffusion}
\begin{figure}[h!]
\begin{center}
\includegraphics[width=10cm]{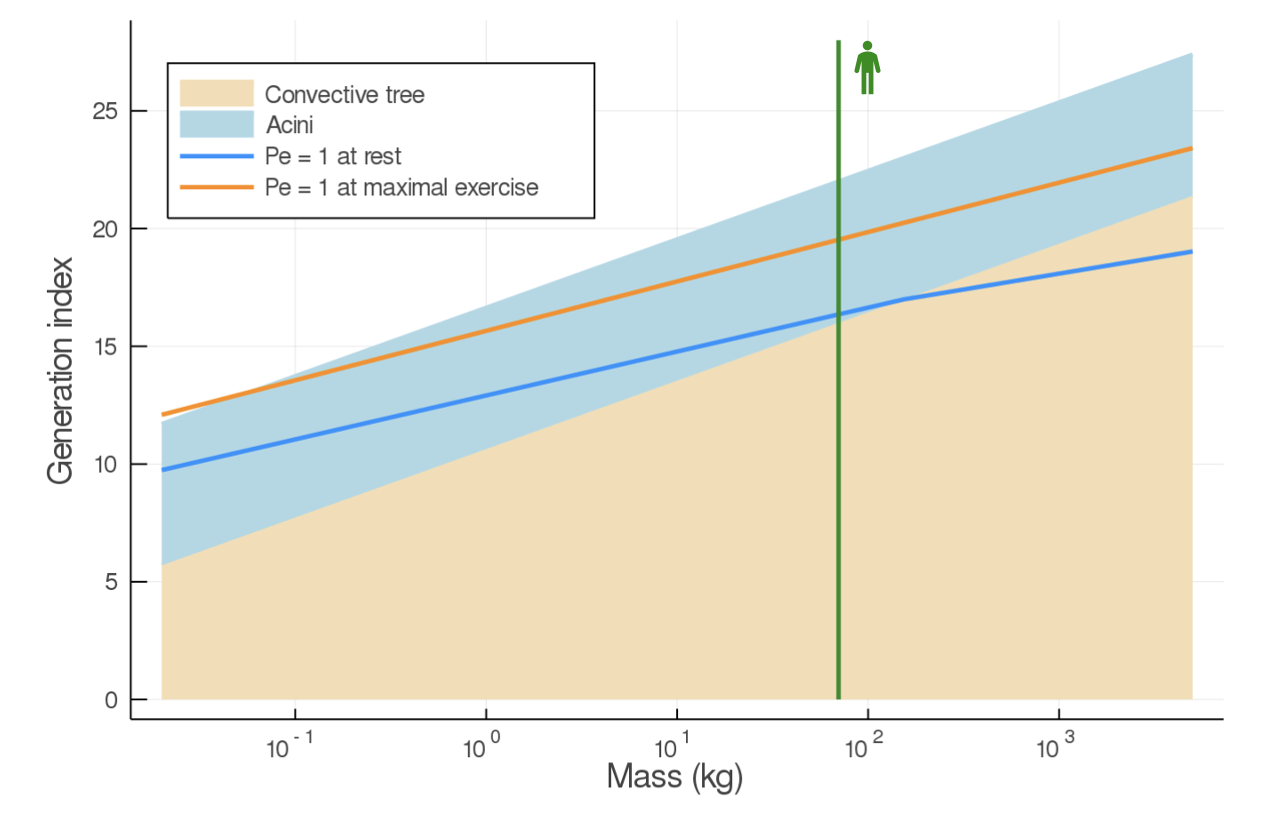}
\caption{Localization of the transition between convective and diffusive transport of oxygen in the lung as a function of animal mass (logarithmic scale). 
The lines correspond to the localizations of that transition at BMR (rest, blue line) and MMR ($\vom$, orange line). 
The vertical green line corresponds to human mass ($70$ kg).
The lower beige region corresponds to the convective zone of the lung and the upper blue region corresponds to the exchange surface (acini). 
Small mammals tend to transport oxygen mainly by convection. 
Hence, there is no screening effect~\parencite{sapoval_smaller_2002} affecting their exchange surface. Their oxygen pressure gradients between alveoli and blood are maximal everywhere, making their pulmonary system very efficient. 
To the contrary, due to the screening effect, large mammals use only a small portion of their exchange surface at rest. 
Hence, large mammals have a large reserve of exchange surface available for higher metabolic rates. 
Increasing the ventilation amplitude decreases the screening effect. 
Hence, the pulmonary system of large mammals is more efficient at exercise than at rest.
}
\label{fig:convDiff}
\end{center}
\end{figure}

The localization of the transition between convective and diffusive transport can be estimated with the P\'{e}clet number~\parencite{noel_interplay_2019}.
This number measures the relative influence of the transport of oxygen by convection on the transport by diffusion. 
In our model, the localization of the transition zone corresponds to the generation $k$, where the Péclet number, denoted $\text{Pe}_k$, becomes smaller than one, see Appendix \ref{fn12}.
The generation index $k$ at which the transition occurs depends on the mammal mass $M$ and on the air flow rate $\ve$ in the mammal lung, see Appendix \ref{fn13},
\begin{equation}
2^k \propto 
\left\{
\begin{array}{ll}
\ve^{\frac32} \times M^{-\frac34} & \text{if $k < G$}\\
\ve \times M^{-\frac{5}{24}} & \text{if $k\geqslant G$}
\end{array}
\right.
\end{equation}
At BMR, our model predicts that the convection--diffusion transition occurs in the convective tree for mammals with a mass larger than about $150$ kg and in the acini for the others.  
The corresponding generation index $k_{\rm{BMR}}$ follows the law $2^{k_{\rm{BMR}}} \propto M^{0.405}$ if the mammal mass is larger than about $150$ kg, and $2^{k_{\rm{BMR}}} \propto M^{0.56}$ otherwise.
At MMR, the convection--diffusion transition always occurs in the acini at the generation index $k_{\rm{MMR}}$, which follows the scaling $2^{k_{\rm{MMR}}} \propto M^{0.63}$. 
Hence, in each lung compartment, the location of the transition depends linearly on the logarithm of the animal mass, see Figure \ref{fig:convDiff} and Appendix \ref{fn14}.

At exercise, the transition occurs deeper in the lung than at rest. 
Mammals with a low mass have a transition that is localized relatively deep in their lung, as shown in Figure \ref{fig:convDiff}. 
In the acini, oxygen is simultaneously displaced along alveolar ducts and captured by the blood that flows in the alveoli walls. 
Consequently, the first alveolar ducts get higher oxygen concentration than those deeper in the acini. 
This phenomenon is known as the screening effect~\parencite{sapoval_smaller_2002} and results in an exchange surface that can be only partially active, depending on the localization in the lung of the transition between convection and diffusion. 
Our model predicts that small mammals are using almost all the volume of their lungs at rest, with low screening effect. 
To the contrary, large mammals present a clear difference in term of volume usage between rest and exercise, with a transition localized near the end of the bronchial tree at rest, implying a strong screening effect, and with a transition localized deeper in the acini at exercise, implying a lower screening effect.

\subsection*{Exhaled oxygen fraction}

The oxygen flow captured by the lung is a proportion of the air flow inhaled, $\vo = \ve \ (f_I - f_E)$ with $\ve = V_T f_b$ the air flow rate, $f_I$ the oxygen fraction in ambient air and $f_E$ the mean exhaled oxygen fraction. The allometric laws predicted by our model for tidal volumes and breathing rates allow to derive similar laws for the drop in oxygen fraction between ambient and exhaled air, $\Delta f = f_I - f_E$: $\Delta f^{\rm{\rm{BMR}}} = 4.61 \ M^{0.002} \ \%$, $\Delta f^{\rm{\rm{FMR}}} = 5.02 \ M^{-0.009} \ \%$ and $\Delta f^{\rm{\rm{MMR}}} = 5.12 \ M^{0.0005} \ \%$. The drop in oxygen fraction depends only slightly on mammal mass and is in the range $3$ to $5 \%$, whatever the ventilation regime. With an inhaled oxygen fraction in air of about $21 \%$, the oxygen fraction in the exhaled air is ranging from $16$ to $18\%$, in full accordance with physiology~\parencite{weibel_pathway_1984}. The quantity $\eta = \Delta f / f_I$ can be considered as a measure for the efficiency of oxygen extraction by the lung. Our model suggests that the system extraction is optimal for values of $\eta$ of about $20 \%$. 
Differences in $\eta$ exists between small and large mammals because of the non zero exponents in the allometric scaling laws of $\Delta f$. However, the values of these exponents are small and cannot be interpreted as such. They might be the results of the simplifications made in the model and/or of the numerical approximations.

\section*{Discussion}

From a set of core morphometric parameters that represent the lung geometry, our model allows to predict, at any metabolic regime, a set of dynamical parameters that represent the lung ventilation and that minimize an estimation of the mechanical cost of ventilation. 
This approach is able to predict with good accuracy the allometric scaling laws of mammals tidal volumes and breathing frequencies reported in the literature (tidal volume at rest, breathing frequencies at rest and $\vom$~\parencite{stahl_scaling_1967, worthington_relationship_1991, altringham_power_1991, young_properties_1992, west_general_1997}). 
The validation of our model at both minimal and maximal metabolic regimes suggests that its predictions should be valid whatever the regime, in the limit of the availability of the input parameters.
This indicates that the mechanical energy spent for ventilation might have driven the selection by evolution of the ventilation patterns. 

The optimization process was constrained, because the lung has to fulfill the function of transporting the needed respiratory gas to and from blood. 
Although our model mimics only the function of transporting oxygen, it is nevertheless able to reach valid predictions.
This raises the question about the influence of other respiratory gases, and more specifically, of carbon dioxide.
To answer this question, we adapted our model to account for a constraint on the carbon dioxide flow, based on \textcite{noel_interplay_2019}. 
At rest, the new predictions were similar to that of a constrained oxygen flow. 
At exercise, we observed a shift between the two cases, with the constraint on oxygen flow leading to better predictions.
This suggests that the oxygen flow could have driven the selection of breathing rates and tidal volumes in mammals.
This result might seem counterintuitive at first, as oxygen flow is suspected to have a low influence on the control of ventilation at intraspecific level~\parencite{robertshaw_mechanisms_2006}.
However, since the input of our model is the characteristic mass of a mammal species, our model is an interspecific model.
As highlighted in the literature~\parencite{witting_general_1997}, interspecific and intraspecific trends can be very different.
Finally, it is important to notice that, apart from the respiratory gas flows, other quantities, not accounted for in our model, are known to affect the control of ventilation such as mechanical, chemical or thermic regulations~\parencite{speakman_heat_2010, dempsey_respiratory_2015, sobac_allometric_2019}, at least at intraspecific level.

The function of respiratory gas transport is dependent on the physical processes on which these transports rely. 
Except for small mammals, the most crucial physical phenomena is the screening effect~\parencite{sapoval_smaller_2002}. 
Screening effect affects how the exchange surface is effectively used and drives at which depth in the lung the convection has to bring oxygen so that diffusion could take over the transport. 
The lung main response to a change in metabolic regime is to adjust the amount of exchange surface actually used.
Hence, only an analysis including a reliable representation of mammal lung and of respiratory gas transport is able to reach predictions compatible with physiology whatever the regime. 

The idealized representation of the bronchial tree and of the exchange surface used in this study accounts for five core characteristics common to all the mammalian lungs, as identified in the literature~\parencite{otis_mechanics_1950, weibel_pathway_1984, west_general_1997, mauroy_optimal_2004, noel_interplay_2019}: a bifurcating tree structure; an homogeneous decrease of the size of bronchi at bifurcations; the trachea size; the alveoli size; and the surface area of the exchange surface. 
At a given metabolic rate, these characteristics are major determinants of the optimal tidal volume and breathing frequency that minimize the energetic cost of ventilation.
This indicates that once the metabolic regime is fixed, the morphology of the lung is probably a core driver of the physiological control of ventilation. 
We tested this hypothesis by altering in our analysis the allometric scaling laws related to the lung geometry. 
We observed corresponding alteration of the predicted laws for tidal volumes and breathing frequencies. 
Since morphology itself has probably been selected by evolution in order to minimize the hydrodynamic resistance in a constrained volume~\parencite{mauroy_optimal_2004, dubois_de_la_sabloniere_shape_2011}, morphology and ventilation patterns are intertwined together in order for the lung to function with a low global energetic cost, i.e. a low hydrodynamic resistance $R$ and a low ventilation cost $\tilde{\P}(V_T,f_b)$, which also depends on $R$.
Actually, this suggests that coevolution of these traits might have occurred in order to keep the cost of breathing as low as possible. 
Our representation of the lung does not account for interspecific differences known to exist between the lungs of mammals, such as different degrees of branching asymmetry, monopodial or bipodial lungs, etc.~\parencite{tawhai_ct-based_2004, mauroy_influence_2010, florens_optimal_2011, monteiro_bronchial_2014}.  
Nevertheless, the predictions of our model for the localization of the convection--diffusion transition in idealized lungs lead to good estimations of the allometric scaling laws for tidal volumes and breathing frequencies, indicating that the morphological parameters included in our model might primarily drive the control of ventilation.

The generation index of the convection--diffusion transition, shown in Figure~\ref{fig:convDiff}, depends linearly on the logarithm of the mammal mass. 
Since the structure of the tree is also governed by allometric scaling laws, the generation index at which the transition between the bronchial tree and the acini occurs also depends linearly on the logarithm of the mammal mass. 
However, the slopes are different and the convection--diffusion transition is located in the acini for small mammals and deep in the bronchial tree for large mammals. 
The reason is that larger mammals actually need less oxygen relatively to their mass than small mammals as $\vo/M \propto M^{-1/4}$ at rest and $\vom/M \propto M^{-1/8}$ at $\vom$. 
Hence, at rest, small mammals use almost their entire exchange surface.
They are subject to a low screening effect, making their lung non limitant, since it is able to respond efficiently to a change in metabolism. 
To the contrary, large mammals tend to use only a small portion of that surface at rest and are subject to a strong screening effect.
Actually, the screened exchange surface in large mammals can be seen as an exchange surface reserve, which can be recruited to allow higher metabolic rates. 
Interestingly, for masses near that of a human, the convection--diffusion transition at rest occurs near the bronchial tree--acini transition~\parencite{sapoval_smaller_2002, karamaoun_new_2018, noel_interplay_2019}.

The ability to increase the metabolic rate plays a crucial role in animal life, for example for foraging or for responding to environmental threat. 
Our model suggests that the proportion of oxygen extracted from ambient air by the lung, found to be about $20 \%$, depends only slightly on metabolic rate. 
More oxygen can be extracted at higher metabolic rates because the volumes of inhaled air are larger. 
Except for small mammals, a larger volume of inhaled air allows to use a larger portion of exchange surface, hence reducing \emph{de facto} the screening effect and accelerating the exchanges speed. 
As a consequence, air has to be renewed at a quicker pace and the breathing rate is increased. 
This last effect is however counterbalanced by the increase of dead volume with the intensity of exercise~\parencite{dempsey_respiratory_2015}. 
However, the increase of dead volume with metabolic rate does not compensate the increase of tidal volume. 
Typically our model predicts that, in humans, the ratio between these two volumes drops from about $40 \%$ at BMR down to about $20 \%$ at MMR, in good agreement with the literature~\parencite{haverkamp_physiologic_2005}. 
Nevertheless, larger dead volumes allow to bring a larger oxygen reserve at the convection--diffusion transition point. 
Hence, relatively lower air renewing rates are needed.
The optimization of the mechanical energy reflects a balance between larger air volume and air renewal rate.
A proper balancing allows to maintain an efficient oxygen diffusion gradient in the acini. 
Our model suggests that this effect plays an important role in the control of breathing rates in small mammals.
It predicts that small mammals should exhibit a breathing rate at MMR that is smaller than at FMR, as shown in Figure \ref{fig:volM}. 
Also, as small mammals exhibit almost no screening effect, the oxygen gradients between alveoli and blood are maximal everywhere in the acini and their lung is very efficient, whatever the regime~\parencite{fregosi_arterial_1984}. 
This efficiency induces an optimal response of the lung to changes in the circulatory parameters and no reserve of exchange surface is needed. 
This brings up the hypothesis that the reserve of exchange surface may compensate the screening effect occurring in the lungs of large mammals.
More specific studies and detailed analyzes of the respiratory system are however needed to confirm or infirm these predictions, in particular studies involving a more realistic coupling with the circulatory system.

Finally, there exists exceptions for which the oxygen demand can exceed the transport capacity of the lung at maximal exercise, such as in human highly trained endurance athletes or in thoroughbred horses~\parencite{dempsey_is_2020, powers_is_2020}.
For these exceptions, the response of the control of ventilation induces increased airways resistances and flow limitations.
As a consequence, the energy cost of ventilation becomes excessive for the metabolism.
Our model could be used to study these configurations and to highlight the biophysical processes of these limitations.

\section*{Conclusion}

Our results highlight the influence of the transport of respiratory gas on the control of ventilation, and more generally, on the behavior of the lung and of the respiratory system. 
Our results contribute to improve our understanding of the allometric scaling of ventilation in mammals. 
They represent a new theoretical framework that highlights the evolution of the respiratory system and its links with the organism metabolism. 
Our work suggests that the dynamical characteristics related to the control of ventilation are highly dependent on the morphological characteristics of the lung. 
This dependence comes from the physical processes involved in oxygen transport. 
Moreover, it has been suggested that several core morphological parameters related to the bronchial tree minimize the hydrodynamic resistance of the lung in a limited volume, so that the exchange surface can fill most of the thoracic space~\parencite{mauroy_optimal_2004, dubois_de_la_sabloniere_shape_2011}. 
Consequently, the control of ventilation is, at least partially, a direct consequence of the repartition of lung space between the bronchial tree and the acini. 
More generally, this highlights the importance of the geometrical constraints in the selection of organs characteristics, not only in terms of morphology, but also in terms of dynamics.

\section*{Supplementary material}
Script and codes are available online \parencite{noel_code_2021}: \href{https://doi.org/10.5281/zenodo.5112934}{doi:10.5281/zenodo.5112934}

\section*{Acknowledgements}

Version 6 of this preprint has been peer-reviewed and recommended by Peer Community In Mathematical and Computational Biology (\href{https://doi.org/10.24072/pci.mcb.100005}{https://doi.org/10.24072/pci.mcb.100005}).

We would like to thank Dr. Elodie Vercken (INRAE, Institut Sophia Agrobiotech, France) for fruitful discussions. This work has been supported by the Agence Nationale de la Recherche, in the frame of the project VirtualChest (ANR-16-CE19-0014) and of the IDEX UCA JEDI (ANR-15-IDEX-01) and by the association Vaincre La Mucoviscidose (RF20190502489).

\section*{Conflict of interest disclosure}

The authors of this article declare that they have no financial conflict of interest with the content of this article.

\printbibliography[notcategory=ignore]

@Article{worthington_relationship_1991,
  author   = {Worthington, J. and Young, I. S. and Altringham, J. D.},
  title    = {The relationship between body mass and ventilation rate in mammals},
  issn     = {0022-0949, 1477-9145},
  language = {en},
  number   = {1},
  pages    = {533--536},
  url      = {https://jeb.biologists.org/content/161/1/533},
  urldate  = {2019-10-08},
  volume   = {161},
  journal  = {Journal of Experimental Biology},
  month    = nov,
  pmid     = {1757778},
  year     = {1991},
}

@Article{dhindsa_comparative_1971,
  author   = {Dhindsa, D. S. and Hoversland, A. S. and Metcalfe, J.},
  title    = {Comparative studies of the respiratory functions of mammalian blood. {VII}. {Armadillo} ({Dasypus} novemcinctus)},
  issn     = {0034-5687},
  language = {eng},
  number   = {2},
  pages    = {198--208},
  volume   = {13},
  journal  = {Respiration Physiology},
  month    = nov,
  pmid     = {5133712},
  year     = {1971},
}

@Article{otis_mechanics_1950,
  author   = {Otis, Arthur B. and Fenn, Wallace O. and Rahn, Hermann},
  title    = {Mechanics of {Breathing} in {Man}},
  doi      = {10.1152/jappl.1950.2.11.592},
  issn     = {8750-7587, 1522-1601},
  language = {en},
  number   = {11},
  pages    = {592--607},
  url      = {http://www.physiology.org/doi/10.1152/jappl.1950.2.11.592},
  urldate  = {2018-12-13},
  volume   = {2},
  journal  = {Journal of Applied Physiology},
  month    = may,
  year     = {1950},
}

@Book{anderson_jr_fundamentals_2010,
  author    = {Anderson Jr, John David},
  title     = {Fundamentals of aerodynamics},
  publisher = {Tata McGraw-Hill Education},
  year      = {2010},
}

@PhdThesis{mauroy_viscosity_2014,
  author     = {Mauroy, Benjamin},
  title      = {Viscosity : an architect for the respiratory system?},
  type       = {Habilitation à diriger des recherches},
  url        = {https://hal.archives-ouvertes.fr/tel-01139846},
  urldate    = {2016-02-29},
  month      = dec,
  school     = {Université de Nice-Sophia Antipolis},
  shorttitle = {Viscosity},
  year       = {2014},
}

@Article{gunther_dimensional_1975,
  author   = {Gunther, B.},
  title    = {Dimensional analysis and theory of biological similarity},
  doi      = {10.1152/physrev.1975.55.4.659},
  issn     = {0031-9333, 1522-1210},
  language = {en},
  number   = {4},
  pages    = {659--699},
  url      = {https://www.physiology.org/doi/10.1152/physrev.1975.55.4.659},
  urldate  = {2019-09-30},
  volume   = {55},
  journal  = {Physiological Reviews},
  month    = oct,
  year     = {1975},
}

@Article{gunther_physiometry_1966,
  author   = {Günther, B. and De la Barra, B. L.},
  title    = {Physiometry of the mammalian circulatory system},
  issn     = {0001-6764},
  language = {eng},
  number   = {1},
  pages    = {32--42},
  volume   = {16},
  journal  = {Acta Physiologica Latino Americana},
  pmid     = {5940180},
  year     = {1966},
}

@Article{monteiro_bronchial_2014,
  author   = {Monteiro, Adilson and Smith, Ricardo Luiz},
  title    = {Bronchial tree {Architecture} in {Mammals} of {Diverse} {Body} {Mass}},
  doi      = {10.4067/S0717-95022014000100050},
  issn     = {0717-9502},
  language = {en},
  number   = {1},
  pages    = {312--316},
  url      = {http://www.scielo.cl/scielo.php?script=sci_arttext&pid=S0717-95022014000100050&lng=en&nrm=iso&tlng=en},
  urldate  = {2018-11-07},
  volume   = {32},
  journal  = {International Journal of Morphology},
  month    = mar,
  year     = {2014},
}

@Article{mauroy_interplay_2003,
  author   = {Mauroy, B. and Filoche, M. and Andrade, J. S. and Sapoval, B.},
  title    = {Interplay between geometry and flow distribution in an airway tree},
  issn     = {0031-9007},
  language = {eng},
  pages    = {148101},
  volume   = {90},
  journal  = {Physical Review Letters},
  month    = apr,
  pmid     = {12731949},
  year     = {2003},
}

@Article{sapoval_smaller_2002,
  author     = {Sapoval, Bernard and Filoche, M. and Weibel, E. R.},
  title      = {Smaller is better—but not too small: {A} physical scale for the design of the mammalian pulmonary acinus},
  doi        = {10.1073/pnas.122352499},
  issn       = {0027-8424, 1091-6490},
  language   = {en},
  number     = {16},
  pages      = {10411--10416},
  url        = {http://www.pnas.org/content/99/16/10411},
  urldate    = {2014-10-24},
  volume     = {99},
  journal    = {PNAS},
  month      = jun,
  pmid       = {12136124},
  shorttitle = {Smaller is better—but not too small},
  year       = {2002},
}

@Article{mauroy_influence_2010,
  author   = {Mauroy, Benjamin and Bokov, Plamen},
  title    = {The influence of variability on the optimal shape of an airway tree branching asymmetrically},
  doi      = {10.1088/1478-3975/7/1/016007},
  issn     = {1478-3975},
  language = {eng},
  number   = {1},
  pages    = {16007},
  volume   = {7},
  journal  = {Phys Biol},
  pmid     = {20090193},
  year     = {2010},
}

@Article{macklem_physiology_1974,
  author   = {MacKlem, P. T.},
  title    = {Physiology of {Cough}},
  doi      = {10.1177/000348947408300611},
  issn     = {0003-4894},
  language = {en},
  note     = {Publisher: SAGE Publications Inc},
  number   = {6},
  pages    = {761--768},
  url      = {https://doi.org/10.1177/000348947408300611},
  urldate  = {2021-02-12},
  volume   = {83},
  journal  = {Ann Otol Rhinol Laryngol},
  month    = nov,
  year     = {1974},
}

@Article{gomes_geometric_2002,
  author   = {Gomes, Rute F.M and Bates, Jason H.T},
  title    = {Geometric determinants of airway resistance in two isomorphic rodent species},
  doi      = {10.1016/S0034-5687(02)00017-8},
  issn     = {15699048},
  language = {en},
  number   = {3},
  pages    = {317--325},
  url      = {https://linkinghub.elsevier.com/retrieve/pii/S0034568702000178},
  urldate  = {2021-06-18},
  volume   = {130},
  journal  = {Respiratory Physiology \& Neurobiology},
  month    = jun,
  year     = {2002},
}

@Book{lodish_molecular_2008,
  author    = {Lodish, Harvey and Darnell, James and Berk, Arnold and Matsudaira, Paul and Kaiser, Chris A. and Krieger, Monty and Scott, Matthew P. and Zipursky, Lawrence},
  title     = {Molecular {Cell} {Biology}},
  edition   = {Fifth Edition},
  publisher = {W. H. Freeman},
  url       = {http://gen.lib.rus.ec/book/index.php?md5=fcea1ade54d39a86c422e3c5ad8d6822},
  year      = {2008},
}

@Article{speakman_heat_2010,
  author   = {Speakman, John R. and Król, Elżbieta},
  title    = {The {Heat} {Dissipation} {Limit} {Theory} and {Evolution} of {Life} {Histories} in {Endotherms}—{Time} to {Dispose} of the {Disposable} {Soma} {Theory}?},
  doi      = {10.1093/icb/icq049},
  issn     = {1557-7023, 1540-7063},
  language = {en},
  number   = {5},
  pages    = {793--807},
  url      = {https://academic.oup.com/icb/article-lookup/doi/10.1093/icb/icq049},
  urldate  = {2019-08-12},
  volume   = {50},
  journal  = {Integrative and Comparative Biology},
  month    = nov,
  year     = {2010},
}

@Book{west_respiratory_2011,
  author     = {West, John B.},
  title      = {Respiratory {Physiology}: {The} {Essentials}},
  edition    = {9th Revised edition},
  isbn       = {978-1-60913-640-6},
  language   = {Anglais},
  publisher  = {Lippincott Williams and Wilkins},
  address    = {Philadelphia},
  month      = aug,
  shorttitle = {Respiratory {Physiology}},
  year       = {2011},
}

@Article{hudson_relationship_2013,
  author   = {Hudson, Lawrence N. and Isaac, Nick J. B. and Reuman, Daniel C.},
  title    = {The relationship between body mass and field metabolic rate among individual birds and mammals},
  doi      = {10.1111/1365-2656.12086},
  issn     = {1365-2656},
  language = {en},
  number   = {5},
  pages    = {1009--1020},
  urldate  = {2020-03-02},
  volume   = {82},
  journal  = {Journal of Animal Ecology},
  year     = {2013},
}

@Article{weibel_allometric_2004,
  author     = {Weibel, Ewald R. and Bacigalupe, Leonardo D. and Schmitt, Beat and Hoppeler, Hans},
  title      = {Allometric scaling of maximal metabolic rate in mammals: muscle aerobic capacity as determinant factor},
  doi        = {10.1016/j.resp.2004.01.006},
  issn       = {1569-9048},
  number     = {2},
  pages      = {115--132},
  url        = {http://www.sciencedirect.com/science/article/pii/S1569904804000126},
  urldate    = {2018-10-15},
  volume     = {140},
  journal    = {Respiratory Physiology \& Neurobiology},
  month      = may,
  shorttitle = {Allometric scaling of maximal metabolic rate in mammals},
  year       = {2004},
}

@Article{dubois_de_la_sabloniere_shape_2011,
  author  = {Dubois de La Sablonière, Xavier and Mauroy, Benjamin and Privat, Yannick},
  title   = {Shape minimization of the dissipated energy in dyadic trees},
  journal = {Discrete and Continuous Dynamical Systems - Series B, American Institute of Mathematical Sciences},
  year    = {2011},
}

@InCollection{agostoni_static_2011,
  author    = {Agostoni, Emilio and Hyatt, Robert E.},
  booktitle = {Comprehensive {Physiology}},
  title     = {Static {Behavior} of the {Respiratory} {System}},
  editor    = {Terjung, Ronald},
  isbn      = {978-0-470-65071-4},
  publisher = {John Wiley \& Sons, Inc.},
  url       = {http://www.comprehensivephysiology.com/WileyCDA/CompPhysArticle/refId-cp030309.html},
  urldate   = {2014-09-25},
  address   = {Hoboken, NJ, USA},
  month     = jan,
  year      = {2011},
}

@Article{tenney_comparative_1967,
  author     = {Tenney, S. M. and Bartlett, D.},
  title      = {Comparative quantitative morphology of the mammalian lung: {Trachea}},
  doi        = {10.1016/0034-5687(67)90002-3},
  issn       = {0034-5687},
  number     = {2},
  pages      = {130--135},
  url        = {http://www.sciencedirect.com/science/article/pii/0034568767900023},
  urldate    = {2016-12-06},
  volume     = {3},
  journal    = {Respiration Physiology},
  month      = oct,
  shorttitle = {Comparative quantitative morphology of the mammalian lung},
  year       = {1967},
}

@Article{white_mammalian_2003,
  author   = {White, Craig R. and Seymour, Roger S.},
  title    = {Mammalian basal metabolic rate is proportional to body mass2/3},
  doi      = {10.1073/pnas.0436428100},
  issn     = {0027-8424},
  number   = {7},
  pages    = {4046--4049},
  url      = {https://www.ncbi.nlm.nih.gov/pmc/articles/PMC153045/},
  urldate  = {2018-06-26},
  volume   = {100},
  journal  = {Proc Natl Acad Sci U S A},
  month    = apr,
  pmcid    = {PMC153045},
  pmid     = {12637681},
  year     = {2003},
}

@Article{bishop_integration_2013,
  author   = {Bishop, Charles M. and Spivey, Robin J.},
  title    = {Integration of exercise response and allometric scaling in endotherms},
  doi      = {10.1016/j.jtbi.2013.01.002},
  issn     = {00225193},
  language = {en},
  pages    = {11--19},
  url      = {https://linkinghub.elsevier.com/retrieve/pii/S0022519313000210},
  urldate  = {2019-09-27},
  volume   = {323},
  journal  = {Journal of Theoretical Biology},
  month    = apr,
  year     = {2013},
}

@PhdThesis{felici_physics_2003,
  author   = {Felici, Maddalena},
  title    = {Physics of the oxygen diffusion in the human lung},
  language = {en},
  month    = jun,
  school   = {Ecole Polytechnique X},
  year     = {2003},
}

@Article{hill_oxygen_1936,
  author  = {Hill, Robert and Wolvekamp, H. P. and Hopkins, Frederick Gowland},
  title   = {The oxygen dissociation curve of haemoglobin in dilute solution},
  number  = {819},
  pages   = {484--495},
  volume  = {120},
  journal = {Proc. R. Soc. Lond. B},
  year    = {1936},
}

@Article{florens_optimal_2011,
  author   = {Florens, Magali and Sapoval, Bernard and Filoche, Marcel},
  title    = {Optimal {Branching} {Asymmetry} of {Hydrodynamic} {Pulsatile} {Trees}},
  doi      = {10.1103/PhysRevLett.106.178104},
  number   = {17},
  pages    = {178104},
  url      = {http://link.aps.org/doi/10.1103/PhysRevLett.106.178104},
  urldate  = {2014-08-25},
  volume   = {106},
  journal  = {Phys. Rev. Lett.},
  month    = apr,
  year     = {2011},
}

@Article{altringham_power_1991,
  author     = {Altringham, J. D. and Young, I. S.},
  title      = {Power output and the frequency of oscillatory work in mammalian diaphragm muscle: the effects of animal size},
  issn       = {0022-0949, 1477-9145},
  language   = {en},
  number     = {1},
  pages      = {381--389},
  url        = {https://jeb.biologists.org/content/157/1/381},
  urldate    = {2019-10-08},
  volume     = {157},
  copyright  = {© 1991 by Company of Biologists},
  journal    = {Journal of Experimental Biology},
  month      = may,
  pmid       = {2061706},
  shorttitle = {Power output and the frequency of oscillatory work in mammalian diaphragm muscle},
  year       = {1991},
}

@Article{johnson_mechanical_1992,
  author   = {Johnson, B. D. and Saupe, K. W. and Dempsey, J. A.},
  title    = {Mechanical constraints on exercise hyperpnea in endurance athletes},
  doi      = {10.1152/jappl.1992.73.3.874},
  issn     = {8750-7587},
  note     = {Publisher: American Physiological Society},
  number   = {3},
  pages    = {874--886},
  url      = {https://journals.physiology.org/doi/abs/10.1152/jappl.1992.73.3.874},
  urldate  = {2020-07-27},
  volume   = {73},
  journal  = {Journal of Applied Physiology},
  month    = sep,
  year     = {1992},
}

@Article{noel_interplay_2019,
  author   = {Noël, Frédérique and Mauroy, Benjamin},
  title    = {Interplay {Between} {Optimal} {Ventilation} and {Gas} {Transport} in a {Model} of the {Human} {Lung}},
  doi      = {10.3389/fphys.2019.00488},
  issn     = {1664-042X},
  language = {English},
  url      = {https://www.frontiersin.org/articles/10.3389/fphys.2019.00488/full?&utm_source=Email_to_authors_&utm_medium=Email&utm_content=T1_11.5e1_author&utm_campaign=Email_publication&field=&journalName=Frontiers_in_Physiology&id=450174},
  urldate  = {2019-04-26},
  volume   = {10},
  journal  = {Front. Physiol.},
  year     = {2019},
}

@Article{mauroy_toward_2015,
  author     = {Mauroy, Benjamin and Flaud, Patrice and Pelca, Dominique and Fausser, Christian and Merckx, Jacques and Mitchell, Barrett R.},
  title      = {Toward the modeling of mucus draining from human lung: role of airways deformation on air-mucus interaction},
  doi        = {10.3389/fphys.2015.00214},
  issn       = {1664-042X},
  language   = {English},
  url        = {https://www.frontiersin.org/articles/10.3389/fphys.2015.00214/full#},
  urldate    = {2017-10-17},
  volume     = {6},
  journal    = {Front. Physiol.},
  shorttitle = {Toward the modeling of mucus draining from human lung},
  year       = {2015},
}

@Article{speakman_history_1998,
  author   = {Speakman, J R},
  title    = {The history and theory of the doubly labeled water technique},
  doi      = {10.1093/ajcn/68.4.932S},
  issn     = {0002-9165, 1938-3207},
  number   = {4},
  pages    = {932S--938S},
  url      = {https://academic.oup.com/ajcn/article/68/4/932S-938S/4648697},
  urldate  = {2019-10-24},
  volume   = {68},
  journal  = {The American Journal of Clinical Nutrition},
  month    = oct,
  year     = {1998},
}

@Article{lindstedt_pulmonary_1984,
  author   = {Lindstedt, S. L.},
  title    = {Pulmonary transit time and diffusing capacity in mammals},
  doi      = {10.1152/ajpregu.1984.246.3.R384},
  issn     = {0363-6119, 1522-1490},
  language = {en},
  number   = {3},
  pages    = {R384--R388},
  url      = {https://www.physiology.org/doi/10.1152/ajpregu.1984.246.3.R384},
  urldate  = {2019-10-07},
  volume   = {246},
  journal  = {American Journal of Physiology-Regulatory, Integrative and Comparative Physiology},
  month    = mar,
  year     = {1984},
}

@Article{stahl_scaling_1967,
  author   = {Stahl, Walter R.},
  title    = {Scaling of respiratory variables in mammals},
  number   = {3},
  pages    = {453--460},
  url      = {http://www.uvm.edu/~pdodds/teaching/courses/2009-08UVM-300/docs/others/1966/stahl1966a.pdf},
  urldate  = {2016-11-17},
  volume   = {22},
  journal  = {J. appl. Physiol},
  year     = {1967},
}

@Article{mauroy_toward_2011,
  author     = {Mauroy, Benjamin and Fausser, Christian and Pelca, Dominique and Merckx, Jacques and Flaud, Patrice},
  title      = {Toward the modeling of mucus draining from the human lung: role of the geometry of the airway tree},
  doi        = {10.1088/1478-3975/8/5/056006},
  issn       = {1478-3975},
  number     = {5},
  pages      = {056006},
  url        = {http://stacks.iop.org/1478-3975/8/i=5/a=056006?key=crossref.4a7ba4c6a63fa409af6c6b3114263ff2},
  urldate    = {2014-07-16},
  volume     = {8},
  journal    = {Physical Biology},
  month      = oct,
  shorttitle = {Toward the modeling of mucus draining from the human lung},
  year       = {2011},
}

@Article{rocco_what_2020,
  author  = {Rocco, Patricia Rieken Macedo and Marini, John J.},
  title   = {What have we learned from animal models of ventilator-induced lung injury?},
  doi     = {10.1007/s00134-020-06143-x},
  issn    = {0342-4642},
  pages   = {1--4},
  url     = {https://www.ncbi.nlm.nih.gov/pmc/articles/PMC7270159/},
  urldate = {2021-06-18},
  journal = {Intensive Care Med},
  month   = jun,
  pmcid   = {PMC7270159},
  pmid    = {32500178},
  year    = {2020},
}

@Article{bezanson_julia_2017,
  author  = {Bezanson, Jeff and Edelman, Alan and Karpinski, Stefan and Shah, Viral B},
  title   = {Julia: {A} fresh approach to numerical computing},
  note    = {Publisher: SIAM},
  number  = {1},
  pages   = {65--98},
  url     = {https://doi.org/10.1137/141000671},
  volume  = {59},
  journal = {SIAM review},
  year    = {2017},
}

@Article{metzger_branching_2008,
  author    = {Metzger, Ross J. and Klein, Ophir D. and Martin, Gail R. and Krasnow, Mark A.},
  title     = {The branching programme of mouse lung development},
  doi       = {10.1038/nature07005},
  issn      = {0028-0836},
  language  = {en},
  number    = {7196},
  pages     = {745--750},
  url       = {http://www.nature.com/nature/journal/v453/n7196/abs/nature07005.html},
  urldate   = {2014-10-19},
  volume    = {453},
  copyright = {© 2008 Nature Publishing Group},
  journal   = {Nature},
  month     = jun,
  year      = {2008},
}

@Book{weibel_pathway_1984,
  author     = {Weibel, Ewald R.},
  title      = {The {Pathway} for {Oxygen}: {Structure} and {Function} in the {Mammalian} {Respiratory} {System}},
  isbn       = {978-0-674-65791-5},
  publisher  = {Harvard University Press},
  shorttitle = {The {Pathway} for {Oxygen}},
  year       = {1984},
}

@Book{witting_general_1997,
  author     = {Witting, Lars},
  title      = {A general theory of evolution: by means of selection by density dependent competitive interactions},
  isbn       = {978-87-90514-00-6},
  language   = {en},
  note       = {OCLC: 39246846},
  publisher  = {Peregrine Publisher},
  address    = {Århus, Denmark},
  shorttitle = {A general theory of evolution},
  year       = {1997},
}

@Book{peters_ecological_1986,
  author    = {Peters, Robert H.},
  title     = {The ecological implications of body size},
  publisher = {Cambridge University Press},
  volume    = {2},
  year      = {1986},
}

@InCollection{haverkamp_physiologic_2005,
  author    = {Haverkamp, Hans C and Dempsey, Jerome A and Miller, Jordan D and Romer, Lee M and Eldridge, Marlowe W},
  booktitle = {Physiologic basis of respiratory disease},
  title     = {Physiologic responses to exercise},
  isbn      = {978-1-55009-236-3},
  pages     = {17},
  publisher = {BC Decker, Inc},
  address   = {Hamilton},
  year      = {2005},
}

@Article{elad_steady_1989,
  author     = {Elad, David and Kamm, Roger D. and Shapiro, Ascher H.},
  title      = {Steady compressible flow in collapsible tubes: application to forced expiration},
  doi        = {10.1017/S0022112089001515},
  issn       = {1469-7645, 0022-1120},
  language   = {en},
  pages      = {401--418},
  url        = {https://www.cambridge.org/core/journals/journal-of-fluid-mechanics/article/steady-compressible-flow-in-collapsible-tubes-application-to-forced-expiration/5F272DE3E4E365897255D4E06C6B10AD},
  urldate    = {2019-03-29},
  volume     = {203},
  journal    = {Journal of Fluid Mechanics},
  month      = jun,
  shorttitle = {Steady compressible flow in collapsible tubes},
  year       = {1989},
}

@Article{kleiber_body_1932,
  author   = {Kleiber, M.},
  title    = {Body size and metabolism},
  issn     = {0073-2230},
  language = {English},
  number   = {11},
  pages    = {315--353},
  url      = {http://hilgardia.ucanr.edu/Abstract/?a=hilg.v06n11p315},
  urldate  = {2018-06-26},
  volume   = {6},
  journal  = {Hilgardia},
  month    = jan,
  year     = {1932},
}

@Article{weibel_exercise-induced_2005,
  author    = {Weibel, Ewald R. and Hoppeler, Hans},
  title     = {Exercise-induced maximal metabolic rate scales with muscle aerobic capacity},
  doi       = {10.1242/jeb.01548},
  issn      = {0022-0949, 1477-9145},
  number    = {9},
  pages     = {1635--1644},
  url       = {http://jeb.biologists.org/content/208/9/1635},
  urldate   = {2019-01-15},
  volume    = {208},
  copyright = {© The Company of Biologists Limited 2005},
  journal   = {Journal of Experimental Biology},
  month     = may,
  pmid      = {15855395},
  year      = {2005},
}

@Article{maina_morphometric_2001,
  author     = {Maina, John N. and van Gils, Peter},
  title      = {Morphometric characterization of the airway and vascular systems of the lung of the domestic pig, {Sus} scrofa: comparison of the airway, arterial and venous systems},
  doi        = {10.1016/S1095-6433(01)00411-1},
  issn       = {10956433},
  language   = {en},
  number     = {4},
  pages      = {781--798},
  url        = {https://linkinghub.elsevier.com/retrieve/pii/S1095643301004111},
  urldate    = {2021-06-18},
  volume     = {130},
  journal    = {Comparative Biochemistry and Physiology Part A: Molecular \& Integrative Physiology},
  month      = nov,
  shorttitle = {Morphometric characterization of the airway and vascular systems of the lung of the domestic pig, {Sus} scrofa},
  year       = {2001},
}

@Article{sobac_allometric_2019,
  author   = {Sobac, Benjamin and Karamaoun, Cyril and Haut, Benoit and Mauroy, Benjamin},
  title    = {Allometric scaling of heat and water exchanges in the mammals' lung},
  note     = {arXiv: 1911.11700},
  url      = {http://arxiv.org/abs/1911.11700},
  urldate  = {2019-12-19},
  journal  = {arXiv:1911.11700 [physics]},
  month    = dec,
  year     = {2019},
}

@Article{west_general_1997,
  author   = {West, Geoffrey B. and Brown, James H. and Enquist, Brian J.},
  title    = {A general model for the origin of allometric scaling laws in biology},
  number   = {5309},
  pages    = {122--126},
  url      = {http://www.sciencemag.org/content/276/5309/122.short},
  urldate  = {2014-07-16},
  volume   = {276},
  journal  = {Science},
  year     = {1997},
}

@Article{rodriguez_pulmonary_1987,
  author     = {Rodriguez, M. and Bur, S. and Favre, A. and Weibel, E. R.},
  title      = {Pulmonary acinus: {Geometry} and morphometry of the peripheral airway system in rat and rabbit},
  doi        = {10.1002/aja.1001800204},
  issn       = {1553-0795},
  number     = {2},
  pages      = {143--155},
  url        = {https://anatomypubs.onlinelibrary.wiley.com/doi/abs/10.1002/aja.1001800204},
  urldate    = {2019-11-07},
  volume     = {180},
  journal    = {American Journal of Anatomy},
  shorttitle = {Pulmonary acinus},
  year       = {1987},
}

@Article{young_properties_1992,
  author    = {Young, I. S. and Warren, R. D. and Altringham, J. D.},
  title     = {Some properties of the mammalian locomotory and respiratory systems in relation to body mass},
  issn      = {0022-0949, 1477-9145},
  language  = {en},
  number    = {1},
  pages     = {283--294},
  url       = {https://jeb.biologists.org/content/164/1/283},
  urldate   = {2019-10-07},
  volume    = {164},
  copyright = {© 1992 by Company of Biologists},
  journal   = {Journal of Experimental Biology},
  month     = mar,
  pmid      = {1583442},
  year      = {1992},
}

@InCollection{dempsey_respiratory_2015,
  author    = {Dempsey, Jerome A. and Jacques, Anthony J.},
  booktitle = {Fishman's {Pulmonary} {Diseases} and {Disorders}},
  title     = {Respiratory {System} {Response} to {Exercise} in {Health}},
  edition   = {5},
  editor    = {Grippi, Michael A. and Elias, Jack A. and Fishman, Jay A. and Kotloff, Robert M. and Pack, Allan I. and Senior, Robert M. and Siegel, Mark D.},
  publisher = {McGraw-Hill Education},
  url       = {accessmedicine.mhmedical.com/content.aspx?aid=1122355615},
  urldate   = {2020-03-06},
  address   = {New York, NY},
  year      = {2015},
}

@Article{agostini_postural_2011,
  author   = {Agostini, Valentina and Chiaramello, Emma and Bredariol, Carla and Cavallini, Chanda and Knaflitz, Marco},
  title    = {Postural control after traumatic brain injury in patients with neuro-ophthalmic deficits},
  doi      = {10.1016/j.gaitpost.2011.05.008},
  issn     = {1879-2219},
  language = {eng},
  number   = {2},
  pages    = {248--253},
  volume   = {34},
  journal  = {Gait \& Posture},
  month    = jun,
  pmid     = {21646021},
  year     = {2011},
}

@Article{mead_control_1960,
  author   = {Mead, Jere},
  title    = {Control of respiratory frequency},
  doi      = {10.1152/jappl.1960.15.3.325},
  issn     = {8750-7587, 1522-1601},
  language = {en},
  number   = {3},
  pages    = {325--336},
  url      = {http://www.physiology.org/doi/10.1152/jappl.1960.15.3.325},
  urldate  = {2018-12-13},
  volume   = {15},
  journal  = {Journal of Applied Physiology},
  month    = may,
  year     = {1960},
}

@Book{weibel_morphometry_1963,
  author    = {Weibel, Ewald R. and Cournand, Andre F. and Richards, Dickinson W.},
  title     = {Morphometry of the {Human} {Lung}},
  edition   = {1 edition},
  isbn      = {978-3-540-03073-7},
  language  = {English},
  publisher = {Springer},
  month     = jan,
  year      = {1963},
}

@Article{dempsey_is_2020,
  author   = {Dempsey, Jerome A. and La Gerche, Andre and Hull, James H.},
  title    = {Is the healthy respiratory system built just right, overbuilt, or underbuilt to meet the demands imposed by exercise?},
  doi      = {10.1152/japplphysiol.00444.2020},
  issn     = {8750-7587, 1522-1601},
  language = {en},
  number   = {6},
  pages    = {1235--1256},
  url      = {https://journals.physiology.org/doi/10.1152/japplphysiol.00444.2020},
  urldate  = {2021-03-04},
  volume   = {129},
  journal  = {Journal of Applied Physiology},
  month    = dec,
  year     = {2020},
}

@TechReport{raabe_tracheobronchial_1976,
  author      = {Raabe, O.G. and Yeh, H.C. and Schum, G.M. and Phalen, R.F.},
  institution = {NM: Lovelace Foundation for Medical Education and Research},
  title       = {Tracheobronchial {Geometry}: {Human}, {Dog}, {Rat}, {Hamster}.},
  address     = {Albuquerque},
  year        = {1976},
}

@Article{fregosi_arterial_1984,
  author   = {Fregosi, R. F. and Dempsey, J. A.},
  title    = {Arterial blood acid-base regulation during exercise in rats},
  doi      = {10.1152/jappl.1984.57.2.396},
  issn     = {8750-7587},
  note     = {Publisher: American Physiological Society},
  number   = {2},
  pages    = {396--402},
  url      = {https://journals.physiology.org/doi/abs/10.1152/jappl.1984.57.2.396},
  urldate  = {2020-07-27},
  volume   = {57},
  journal  = {Journal of Applied Physiology},
  month    = aug,
  year     = {1984},
}

@Article{robertshaw_mechanisms_2006,
  author   = {Robertshaw, David},
  title    = {Mechanisms for the control of respiratory evaporative heat loss in panting animals},
  number   = {2},
  pages    = {664--668},
  volume   = {101},
  journal  = {Journal of Applied Physiology},
  year     = {2006},
}

@Article{powers_is_2020,
  author   = {Powers, Scott K.},
  title    = {Is the lung built for exercise?},
  doi      = {10.1152/japplphysiol.00819.2020},
  issn     = {1522-1601},
  language = {eng},
  number   = {6},
  pages    = {1233--1234},
  volume   = {129},
  journal  = {J Appl Physiol (1985)},
  month    = dec,
  pmid     = {33119466},
  year     = {2020},
}

@Article{tenney_quantitative_1970,
  author     = {Tenney, S. M. and Tenney, J. B.},
  title      = {Quantitative morphology of cold-blooded lungs: {Amphibia} and reptilia},
  doi        = {10.1016/0034-5687(70)90071-X},
  issn       = {0034-5687},
  number     = {2},
  pages      = {197--215},
  url        = {http://www.sciencedirect.com/science/article/pii/003456877090071X},
  urldate    = {2018-08-30},
  volume     = {9},
  journal    = {Respiration Physiology},
  month      = may,
  shorttitle = {Quantitative morphology of cold-blooded lungs},
  year       = {1970},
}

@Book{johnson_biomechanics_2007,
  author     = {Johnson, Arthur T.},
  title      = {Biomechanics and {Exercise} {Physiology}: {Quantitative} {Modeling}},
  isbn       = {978-1-4200-1907-0},
  note       = {Google-Books-ID: oIvMBQAAQBAJ},
  publisher  = {CRC Press},
  month      = mar,
  shorttitle = {Biomechanics and {Exercise} {Physiology}},
  year       = {2007},
}

@Article{karamaoun_new_2018,
  author   = {Karamaoun, Cyril and Sobac, Benjamin and Mauroy, Benjamin and Muylem, Alain Van and Haut, Benoît},
  title    = {New insights into the mechanisms controlling the bronchial mucus balance},
  doi      = {10.1371/journal.pone.0199319},
  issn     = {1932-6203},
  language = {en},
  number   = {6},
  pages    = {e0199319},
  url      = {http://journals.plos.org/plosone/article?id=10.1371/journal.pone.0199319},
  urldate  = {2018-06-23},
  volume   = {13},
  journal  = {PLOS ONE},
  month    = jun,
  year     = {2018},
}

@Article{mauroy_optimal_2004,
  author   = {Mauroy, B. and Filoche, M. and Weibel, E. R. and Sapoval, B.},
  title    = {An optimal bronchial tree may be dangerous},
  doi      = {10.1038/nature02287},
  issn     = {1476-4687},
  language = {eng},
  number   = {6975},
  pages    = {633--636},
  volume   = {427},
  journal  = {Nature},
  month    = feb,
  pmid     = {14961120},
  year     = {2004},
}

@InCollection{terjung_lung_2016,
  author    = {Hsia, Connie C.W. and Hyde, Dallas M. and Weibel, Ewald R.},
  booktitle = {Comprehensive {Physiology}},
  title     = {Lung {Structure} and the {Intrinsic} {Challenges} of {Gas} {Exchange}},
  doi       = {10.1002/cphy.c150028},
  editor    = {Terjung, Ronald},
  isbn      = {978-0-470-65071-4},
  language  = {en},
  pages     = {827--895},
  publisher = {John Wiley \& Sons, Inc.},
  url       = {http://doi.wiley.com/10.1002/cphy.c150028},
  urldate   = {2019-09-27},
  address   = {Hoboken, NJ, USA},
  month     = mar,
  year      = {2016},
}

@Article{henke_regulation_1988,
  author   = {Henke, K. G. and Sharratt, M. and Pegelow, D. and Dempsey, J. A.},
  title    = {Regulation of end-expiratory lung volume during exercise},
  doi      = {10.1152/jappl.1988.64.1.135},
  issn     = {8750-7587},
  note     = {Publisher: American Physiological Society},
  number   = {1},
  pages    = {135--146},
  url      = {https://journals.physiology.org/doi/abs/10.1152/jappl.1988.64.1.135},
  urldate  = {2020-07-27},
  volume   = {64},
  journal  = {Journal of Applied Physiology},
  month    = jan,
  year     = {1988},
}

@Article{huxley_terminology_1936,
  author    = {Huxley, J. S. and Teissier, G.},
  title     = {Terminology of {Relative} {Growth}},
  doi       = {10.1038/137780b0},
  issn      = {1476-4687},
  language  = {eng},
  number    = {3471},
  pages     = {780--781},
  urldate   = {2019-10-08},
  volume    = {137},
  copyright = {1936 Nature Publishing Group},
  journal   = {Nature},
  month     = may,
  year      = {1936},
}

@Article{matute-bello_animal_2008,
  author   = {Matute-Bello, Gustavo and Frevert, Charles W. and Martin, Thomas R.},
  title    = {Animal models of acute lung injury},
  doi      = {10.1152/ajplung.00010.2008},
  issn     = {1040-0605},
  note     = {Publisher: American Physiological Society},
  number   = {3},
  pages    = {L379--L399},
  url      = {https://journals.physiology.org/doi/full/10.1152/ajplung.00010.2008},
  urldate  = {2021-06-18},
  volume   = {295},
  journal  = {American Journal of Physiology-Lung Cellular and Molecular Physiology},
  month    = sep,
  year     = {2008},
}

@Article{tawhai_ct-based_2004,
  author   = {Tawhai, Merryn H. and Hunter, Peter and Tschirren, Juerg and Reinhardt, Joseph and McLennan, Geoffrey and Hoffman, Eric A.},
  title    = {{CT}-based geometry analysis and finite element models of the human and ovine bronchial tree},
  doi      = {10.1152/japplphysiol.00520.2004},
  issn     = {8750-7587},
  language = {eng},
  number   = {6},
  pages    = {2310--2321},
  volume   = {97},
  journal  = {J. Appl. Physiol.},
  month    = dec,
  pmid     = {15322064},
  year     = {2004},
}

@Article{haefeli-bleuer_morphometry_1988,
  author    = {Haefeli-Bleuer, Beatrice and Weibel, Ewald R.},
  title     = {Morphometry of the human pulmonary acinus},
  doi       = {10.1002/ar.1092200410},
  issn      = {1097-0185},
  language  = {en},
  number    = {4},
  pages     = {401--414},
  url       = {http://onlinelibrary.wiley.com/doi/10.1002/ar.1092200410/abstract},
  urldate   = {2014-09-25},
  volume    = {220},
  copyright = {Copyright © 1988 Wiley-Liss, Inc.},
  journal   = {Anat. Rec.},
  month     = apr,
  year      = {1988},
}

@misc{noel_code_2021,
        title = {Code for {The} origin of the allometric scaling of lung's ventilation in mammals},
        url = {https://zenodo.org/record/5112934},
        urldate = {2021-08-31},
        publisher = {Zenodo},
        author = {Noël, Frédérique and Karamaoun, Cyril and Dempsey, Jerome A. and Mauroy, Benjamin},
        month = jul,
        year = {2021},
        doi = {10.5281/zenodo.5112934},
}

\newpage

\begin{center}
\bf \huge Appendix
\end{center}

\hspace{2cm}

\section{Strategy and model hypotheses}
\label{VII}

Table \ref{generalApp} indicates the methodology used in our analysis.
Tables \ref{model1Hyp} and \ref{model2Hyp} on the next pages describe the hypotheses of the two models coupled in our work.

\begin{table*}[h!]
\begin{adjustwidth}{-1.5cm}{}
\footnotesize
\begin{tabular}{lp{0.6\textwidth}l}
\multicolumn{3}{c}{{\bf Biological hypotheses} \parencite{otis_mechanics_1950, mead_control_1960, johnson_biomechanics_2007, noel_interplay_2019}}\\
 \hline
 Evolutive hypothesis & \multicolumn{1}{p{0.01\textwidth}}{} & \multicolumn{1}{p{0.79\textwidth}}{We assume that, in mammals, the ventilation parameters minimize the mechanical power of the ventilation.}\\
 \hline
 Physiological constraint & \multicolumn{1}{p{0.01\textwidth}}{} & \multicolumn{1}{p{0.79\textwidth}}{We focus on the oxygen transport function of the lung and assume that the oxygen flow to the blood has to fit the metabolic regime.}\\
 \hline
 Ventilation parameters & \multicolumn{1}{p{0.01\textwidth}}{} & \multicolumn{1}{p{0.79\textwidth}}{We characterize the ventilation with the breathing frequency $f_b$ and the tidal volume $V_T$} \\
 \hline
\multicolumn{3}{c}{}\\
\multicolumn{3}{c}{\bf Strategy}\\
 \hline
\multicolumn{1}{p{0.2\textwidth}}{\vspace{0.2cm}Our analysis is based on\newline {\bf two input parameters}}&\multicolumn{1}{p{0.01\textwidth}}{}& \multicolumn{1}{p{0.79\textwidth}}{ 
 \begin{itemize}
 \item the mammal mass $\mathbf{M}$
 \item the metabolic need in term of oxygen flow $\mathbf{\vo}$, see Table \ref{metabo}.
 \end{itemize}}\\
 \hline
  \multicolumn{1}{p{0.2\textwidth}}{\vspace{0.5cm}Oxygen flows at typical metabolic rates}&\multicolumn{1}{p{0.01\textwidth}}{}& \multicolumn{1}{p{0.79\textwidth}}{
The amount of oxygen flow needed by the metabolism follows allometric scaling laws that depend on the regime considered.
\begin{itemize}
\item Basal Metabolic Rate (BMR): $\vob \propto M^{\frac34}$~\parencite{kleiber_body_1932, peters_ecological_1986}
\item Field Metabolic Rate (FMR): $\vof \propto M^{0.64}$~\parencite{hudson_relationship_2013}
\item Maximal Metabolic Rate (MMR): $\vom \propto M^{\frac78}$~\parencite{weibel_exercise-induced_2005}
\end{itemize}
 }\\
 \hline
 \multicolumn{1}{p{0.2\textwidth}}{{\bf Two mathematical models} are used to compute estimations of physiological quantities}&\multicolumn{1}{p{0.01\textwidth}}{}&\multicolumn{1}{p{0.79\textwidth}}{
\indent See details in Tables \ref{model1Hyp} and \ref{model2Hyp}, the models inputs are $M$, $\vo$ and the ventilation parameters $\mathbf{f_b}$ and $\mathbf{V_T}$
\begin{itemize}
\item the model 1 estimates the mechanical power to perform the lung ventilation $\mathbf{\tilde{\mathcal{P}}_v(V_T, f_b)}$
\item the model 2 estimates the oxygen flow from lung to blood $\mathbf{f_{O_2}(V_T, f_b)}$
\end{itemize}
}\\
\hline
\multicolumn{1}{p{0.2\textwidth}}{The two models are used for a {\bf constrained optimization process}}&\multicolumn{1}{p{0.01\textwidth}}{}& \multicolumn{1}{p{0.79\textwidth}}{\vspace{-0.1cm}We search for the ventilation parameters $\mathbf{V_T}$ and $\mathbf{f_b}$ that minimize the mechanical power $\mathbf{\tilde{\mathcal{P}}_v(V_T, f_b)}$ with the constraint $\mathbf{f_{O_2}(V_T, f_b) = \vo}$ on the oxygen flow.}\\
 \hline
  \end{tabular}
 \caption{General strategy and hypotheses.}
 \label{generalApp}
 \end{adjustwidth}
\end{table*}
 
 \begin{table*}[h!]
 \begin{adjustwidth}{-3cm}{}
 \footnotesize
\begin{tabular}{lp{0.6\textwidth}l}
 \multicolumn{3}{c}{{\bf Model 1: Power spent for lung ventilation}, adapted from \textcite{otis_mechanics_1950, mead_control_1960, johnson_biomechanics_2007, noel_interplay_2019}}\\
 \hline 
 \multicolumn{3}{c}{The mechanical power spent by lung ventilation has two main sources:}\\
 \multicolumn{3}{c}{the air viscous dissipation in airways and the elastic power stored in thorax tissues}\\
 \hline
 \hline
 {\bf Model inputs:}&\multicolumn{2}{l}{Mammal mass $M$, tidal volume $V_T$ and breathing rate $f_b$}\\
 {\bf Model output:}&\multicolumn{2}{l}{Mechanical power spent by ventilation $\tilde{\mathcal{P}}_v(V_T, f_b)$}\\
 \hline
 \hline
 Viscous dissipation & \multicolumn{2}{p{0.8\textwidth}}{Air viscous dissipation in airways is estimated based on the hydrodynamic resistance of the lung, $R \propto M^{-\frac34}$, see Table \ref{tab:allo_expo}, Appendix \ref{fn3} and \parencite{stahl_scaling_1967}.}\\
 \hline
 Elastic power	 & \multicolumn{2}{p{0.8\textwidth}}{Elastic properties of thorax and lung are estimated based on the lung compliance, $C \propto M^{1}$, see table \ref{tab:allo_expo}, Appendix \ref{fn2} and \parencite{stahl_scaling_1967}.}\\
 \hline
 \end{tabular}
 \caption{Hypotheses of the model 1 that estimates the power spent for ventilating the lung.}
 \label{model1Hyp}
 \end{adjustwidth}
\end{table*}
\newpage
\FloatBarrier
\begin{table*}[h!]
 \begin{adjustwidth}{-3cm}{}
 \footnotesize
\begin{tabular}{lp{0.6\textwidth}l} 
 \multicolumn{3}{c}{{\bf Model 2: Oxygen transport in the lung}, adapted from \textcite{noel_interplay_2019}}\\
 \hline
 \multicolumn{3}{c}{Oxygen is transported in airways by convection with air and by diffusion.}\\ 
 \multicolumn{3}{c}{In acini, oxygen is also exchanged with blood through the airways wall.}\\
 \hline
 \hline
 {\bf Model inputs:}&\multicolumn{2}{l}{Mammal mass $M$, tidal volume $V_T$ and breathing rate $f_b$}\\
 {\bf Model output:}&\multicolumn{2}{l}{Oxygen flow to blood $f_{O_2}(V_T, f_b)$}\\
 \hline
 \hline
  \multicolumn{1}{p{0.2\textwidth}}{\vspace{1cm} Lung geometry} & \multicolumn{2}{p{0.8\textwidth}}{
  The topology of the geometrical model for the mammal lung is based on the literature ~\parencite{weibel_pathway_1984, mauroy_optimal_2004}.
  \begin{itemize}
  \item The lung is modeled as a bifurcating tree, where each airway is a cylinder.
  \item The tree consists in two regions, a conducting zone and a respiratory zone.
  \item The geometry of bifurcations are neglected.
  \end{itemize}
  The tree is scaled using scaling laws for mammals from the literature.
  \begin{itemize}
  \item The root of the tree has a radius $r_0 \propto M^{\frac38}$ ~\parencite{west_general_1997} and a length $l_0 \propto M^{\frac14}$, see Appendix \ref{fn6}, the prefactor of $r_0$ accounts for the dependence of dead volume on metabolic regime, see Appendix \ref{fn5} and~\parencite{johnson_mechanical_1992, dempsey_respiratory_2015}.
  \item The size of the airways decreases at each bifurcation with a constant ratio $h = \left(\frac12\right)^{\frac13}$  in the conducting zone~\parencite{west_general_1997, mauroy_optimal_2004} and remains the same in the respiratory zone, see Figure \ref{fig:tree}.
  \item The conductive zone ends at the generation index $G$ when the radius of the smallest conductive airway reaches that of the alveoli radius, $r_A \propto M^{\frac{1}{12}}$, see Appendix \ref{fn7}.
  \item The number of generations $H$ in the respiratory zone is assumed independent on animal mass and equal to $6$~\parencite{rodriguez_pulmonary_1987, haefeli-bleuer_morphometry_1988}.
  \item The amount $\rho_S$ of exchange surface area per unit of wall surface area of airway in the respiratory zone is determined based on the allometric scaling law of the exchange surface $S_A \propto M^{\frac{11}{12}}$, see Appendix \ref{fn8} and \textcite{west_general_1997}. 
  \end{itemize}
  }\vspace{-0.3cm}\\
 \hline
 \multicolumn{1}{p{0.2\textwidth}}{\vspace{-0.1cm} Air fluid dynamics} & \multicolumn{2}{p{0.8\textwidth}}
 {
 Our model uses the mean air velocity in airways and accounts for the air flow conservation at each bifurcation ~\parencite{mauroy_optimal_2004, noel_interplay_2019}.
 }\\
 \hline
 \multicolumn{1}{p{0.2\textwidth}}{\vspace{-0.1cm} Oxygen transport} & \multicolumn{2}{p{0.8\textwidth}}
 {Oxygen transport occurs by convection with air and by diffusion, see Appendix \ref{I} ~\parencite{noel_interplay_2019}}\\
 \hline
 \multicolumn{1}{p{0.2\textwidth}}{\vspace{0.5cm} Oxygen exchange\newline with blood} & \multicolumn{2}{p{0.8\textwidth}}
 {
 As in \textcite{noel_interplay_2019}, the physics of the oxygen exchange between alveolar air and blood is based on a diffusion process through a membrane.
 \begin{itemize}
 \item The physical properties of the alveolar--capillary membrane is assumed to be equivalent to that of a water membrane.
 \item The thickness of the alveolar--capillary membrane $\tau$ is assumed independent of mass, $\tau \simeq 1 \ \mu \text{m}$~\parencite{sapoval_smaller_2002}.
 \item The flow of oxygen through the membrane is assumed equal to the flow of oxygen stored by the blood flowing in the capillaries~\parencite{felici_physics_2003, noel_interplay_2019}, see Appendix \ref{II}.
 \item The blood flow follows an allometric scaling law based on the transit time of blood in capillaries, see Appendix \ref{II}.
 \end{itemize}
 }\vspace{-0.3cm}\\
 \hline
 \end{tabular}
 \caption{Hypotheses of the model 2 that simulates the transport of oxygen in the lung.}
 \label{model2Hyp}
 \end{adjustwidth}
\end{table*}
\FloatBarrier


\section{Details of the model computations}
\label{VIII}

\subsection{Tidal volume}
\label{fn1}

Tidal volume is computed as the integral of the air flow $u(t) S_0$ over a half ventilation cycle, with $u(t)$ the sine function defined in equation (\ref{velocity}),

$$V_T = \protect\int_0^{\frac{T}2} S_0 u(t) \text{d}t = \frac{U S_0 T}{\pi}$$
As $f_b = 1/T$, the parameterization is equivalent for $(U, T)$ and $(V_T, f_b)$.

\subsection{Power associated to the compliance of the lung}
\label{fn2}

The lung compliance is estimated by the ratio between the shift $V$ in lung volume from functional residual capacity (FRC) and the corresponding shift in pleural pressure $p_{pl}$.
The elastic energy stored is then $\mathcal{E}_e = \frac 12 p_{pl} V = \frac12 \frac{V^2}{C}$.
Finally, the instantaneous elastic power is $\frac{d\mathcal{E}}{dt} = \frac1C V(t) \frac{dV}{dt}$ with $V(t) = \protect\int_0^t S_0 u(\xi) d\xi$. 
We recall that $u(t)$ is a sine function, see equation (\ref{velocity}).
Assuming that the elastic power is stored during inspiration only, its averaged value over a ventilation cycle is
$$
\mathcal{P}_e(U,T) = \frac1T \protect\int_0^{\frac{T}2}\frac1C V(t) \frac{dV}{dt}(t) \text{d}t = \frac1C \frac{U^2 S_0^2 T}{2 \pi^2}
$$
Using the variables $f_b$ and $V_T$ leads to $\tilde{\mathcal{P}}_e(V_T,f_b) = \frac{V_T^2 f_b}{2 C}$.

\subsection{Power associated to the hydrodynamic resistance of the lung}
\label{fn3}

The hydrodynamic resistance of the airway tree $R$ is the ratio between the air pressure drop $\Delta p$ applied between the root and the leaves of the tree and the resulting total air flow going through that tree $\Phi$.
The instantaneous power relative to the viscous dissipation in the airway tree is then $\Delta p \ \Phi = R \Phi^2$.
In our model, $\Phi(t) = u(t) S_0$ with $u(t)$ the sine function defined in equation (\ref{velocity}).
Finally, we average the instantaneous power over a ventilation cycle assuming that the power is spent only during inspiration, 
$$
\mathcal{P}_v(U,T) = \frac1T \protect\int_0^{\frac{T}2} R (u(t) S_0)^2 \text{d}t = R \frac{U^2 S_0^2}4
$$
Using the variables $f_b$ and $V_T$ leads to $\tilde{\mathcal{P}}_v(V_T,f_b) = (\pi V_T f_b)^2 R / 4$.

\subsection{Airway surface area and velocity versus generation index}
\label{fn4}
The cross-section surface area of a branch in the generation $i$ is $S_i = \pi r_i^2$. 
In the model of the bronchial tree ($i=0...G-1$), $S_i = h^{2i} S_0$, while in the model of the acini ($i=G...N-1$), $S_i = S_{G-1}$. 
Air is assumed incompressible in the lung under normal ventilation conditions ~\parencite{elad_steady_1989}, except perhaps during cough~\parencite{macklem_physiology_1974}.
This hypothesis is justified by the value of the air Mach number $\rm{Ma}$ in the lung.
This number is computed with $\rm{Ma} = U/c$, where $U$ is the maximal velocity in airways --reached in the trachea--, and $c$ the speed of sound in air.
The speed of sound in air is $c = \sqrt{1.4 P / \rho}$ with $\rho \simeq 1.2$ kg.m$^{-3}$ the density of air and $P$ the absolute pressure in the lung that can be considered in the range $1000 \pm 100$ cmH$_2$O at the different regimes studied in this work.
Hence, for air velocities $U$ below $100$ m.s$^{-1}$ (or air flow in the lung below $25$ L/s), the Mach number remains below $0.3$, which is a typical threshold under which compressible effects can be neglected~\parencite{anderson_jr_fundamentals_2010}.
Consequently, flow conservation leads to
$$
u_i(t) = \left\{
\begin{array}{ll}
u(t) \left(\frac{1}{2 h^2}\right)^i  &\text{ for } i=0...G-1\\
u_{G-1}(t) \left(\frac{1}{2}\right)^{i-G+1}  &\text{ for } i=G...N-1
\end{array}
\right.
$$

\subsection{Tracheal radius}
\label{fn5}

From \parencite{west_general_1997}, the tracheal radius scales as $r_0 = a M^{\frac38}$.
The prefactor $a$ depends on metabolic rate and is determined based on human data and dead volumes: $a = 1.83 \ 10^{-3}$ m.kg$^{-\frac{3}8}$ at BMR, $a = 1.93 \ 10^{-3}$ m.kg$^{-\frac{3}8}$ at FMR and $a = 2.34 \ 10^{-3}$ m.kg$^{-\frac{3}8}$ at MMR.

\subsection{Tracheal length allometric scaling law}
\label{fn6}
\label{l0comp}

In our model, dead volume is proportional to tracheal volume and $V_{\text{dead}} \propto M^1$~\parencite{tenney_comparative_1967}.
Then, $V_{\text{dead}} \propto \pi r_0^2 l_0 \propto M^1$ leads to $l_0 \propto M^{\frac14}$.

\subsection{Conductive airway generations}
\label{fn7}

The computation of $G$ is based on the hypothesis that the radius of the alveolar ducts are similar to the radius $r_A$ of the alveoli~\parencite{weibel_pathway_1984}, for which an allometric scaling law is known, $r_A \propto M^{\frac1{12}}$~\parencite{west_general_1997}. 
Then, the number of generations $G$ of the bronchial tree is obtained from $r_A = r_{G-1} = r_0 h^{G-1}$, and the number of terminal bronchioles follows
$$
2^{G-1} \propto M^{\frac78}
$$ 
This last allometric scaling law can be rewritten in the form $$G = \left[\frac{\log(r_A/r_0)}{\log(h)} \right] + 1 = \left[\frac78\frac{\log(M)}{\log(2)} + \text{cst} \right] + 1$$

\subsection{Total gas exchange surface of the lung}
\label{fn8}

The total gas exchange surface of the lung $S_A \propto M^{\frac{11}{12}}$~\parencite{west_general_1997} is distributed over the alveolar ducts walls. 
In our model, a single alveolar duct has a lateral surface $s_{\rm{ad}} = 2 \pi r_A l_A$ with $l_A = l_0 h^{G-1} \propto M^{-\frac1{24}}$, hence $s_{\rm{ad}} \propto M^{\frac1{24}}$. 
The total surface of alveolar ducts in the idealized lung is then 
$$S_{\rm{ad}} = 2^{G} \protect\sum_{k=0}^{H-1} 2^k s_{\rm{ad}} = 2^{G} (2^{H}-1) s_{ad} \propto M^{\frac{11}{12}}$$ 
Hence, the amount of exchange surface area per unit of alveolar duct surface area, $\rho_s = S_A/S_{\rm{ad}}$ is such that the product $\rho_s (2^H-1) \propto M^0$ is independent of the animal mass. 
The number of generations of alveolar ducts in an acinus is considered independent of the mass~\parencite{rodriguez_pulmonary_1987, haefeli-bleuer_morphometry_1988}.
Consequently, in our model $\rho_s$ is also independent of the mass. 
Under these conditions, our model respects the allometric scaling law from the literature $S_A \propto M^{\frac{11}{12}}$.

\subsection{Flow rate of oxygen partial pressure per unit length of alveolar ducts}
\label{fn9}

The thickness of the alveolar--capillary membrane $\tau$ is assumed independent of the mass, $\tau \simeq 1 \ \mu \text{m}$~\parencite{sapoval_smaller_2002}. 
The diffusivity $D_{O_2,H_2O}$ of oxygen in tissues can be approximated by its value in water~\parencite{sapoval_smaller_2002}.
The flow rate of oxygen partial pressure per unit length of an alveolar duct is then

\begin{align*}
\beta_i \left(P_{i} - P_{\text{blood}} \right)
= &\ \rho_s \frac{2 \pi r_A}{\pi r_A^2} \kappa \sigma_{O_2,H_2O} \frac{D_{O_2,H_2O} }{\tau} \left(P_{i} - P_{\text{blood}} \right)\\ 
= &\ \rho_s \frac{2\kappa}{r_A} \alpha\left(P_{i} - P_{\text{blood}} \right)
\end{align*} 
where $\kappa$ is the ratio relating partial pressure of the gas to its concentration in water, 
$\sigma_{O_2,H_2O}$ is the solubility coefficient 
of the gas in water and $D_{O_2,H_2O}$ is the diffusion coefficient 
of the gas in water. 
The permeability of the alveolar membrane $\alpha$ is $\alpha = \sigma_{O_2,H_2O} \frac{D_{O_2,H_2O} }{\tau}$.

\subsection{Total flow of oxygen exchanged with the blood $f_{O_2}(V_T, f_b)$}
\label{fn10}

The estimation of the total flow of oxygen exchanged with blood $f_{O_2}(V_T, f_b)$ is computed from an established ventilation cycle, 
\begin{equation}
f_{O_2}(V_T, f_b) =
\frac{2 \pi r_A \alpha \rho_s}{T}\protect\sum \limits_{i=G}^{N-1} 2^i \protect\int_{t_C}^{t_C+T} \protect\int_{0}^{l_i}   \left(P_{i}(t,x)-P_{\text{blood}}(t,x)\right) \text{d}x \, \text{d}t
\end{equation}
with $t_C$ a time at which the system has reached a periodic regime and $T = 1/f_b$.

\subsection{Otis et al. optimal breathing frequency at rest}
\label{fn11}

The optimal breathing frequency computed by Otis et al. was obtained by canceling the derivative of the power relatively to $f_b$~\parencite{otis_mechanics_1950, johnson_biomechanics_2007},
\begin{equation}
f_{b,\rm{pred}} = \frac{2 \va/V_D}{1 + \sqrt{1+4 \pi^2 R C \va / V_D}}.
\end{equation}
At BMR, the allometric scaling laws of all the physiological quantities involved in this expression for $f_b$ are available in the literature: $\va \propto M^{\frac34}$~\parencite{gunther_dimensional_1975}, $V_D \propto M^1$~\parencite{stahl_scaling_1967}, $R \propto M^{-\frac34}$~\parencite{stahl_scaling_1967, west_general_1997} and $C \propto M^1$~\parencite{stahl_scaling_1967}. Hence, we are able to derive an allometric scaling law for breathing rate at BMR, $f_{b}^{\rm{\rm{BMR}}}$, based on ventilation data in healthy young humans~\parencite{haverkamp_physiologic_2005},
$$
f_{b,\rm{pred}}^{\rm{\rm{BMR}}} = 0.9 \ M^{-\frac14} \ \rm{Hz} 
$$
Based on breathing frequency and on ventilation data from \textcite{haverkamp_physiologic_2005}, we can deduce the allometric scaling law for tidal volumes at BMR, $V_{T}^{\rm{\rm{BMR}}} = \va / f_{b}^{\rm{\rm{BMR}}} + V_D$. Since  $\va / f_{b}^{\rm{\rm{BMR}}} \propto M^{\frac34} / M^{-\frac14}$ and $V_D \propto M^1$, we have 
$$
V_{T,\rm{pred}}^{\rm{\rm{BMR}}} =  7.5 \ M^1 \ \rm{ml} 
$$

\subsection{Péclet number}
\label{fn12}

The Péclet number is computed by rewriting the transport equations (\ref{eq:eq_1}) in a dimensionless form,
\begin{equation}
\frac{2 l_i^2}{D T}\frac{\partial P_{i}}{\partial s} - \frac{\partial ^2 P_{i}}{\partial \xi^2} + \underbrace{\frac{l_i u_i(sT/2)}{D}}_{Pe_i(s)}\frac{\partial P_{i}}{\partial \xi} + \frac{\beta_i l_i^2}{D} \left(P_{i} - P_{\text{blood}} \right) = 0, \\
\text{ for } \xi \in [0,1]
 \label{eq:eq_x}
\end{equation}
The dimensionless time is $s = 2t/T$ with $T/2$ the inspiration or expiration time and the dimensionless space is $\xi = x/l_i$. 
We define $\rm{Pe}_i$ as the average of the time-dependent P\'{e}clet number $Pe_i(s)$ over a half breath cycle.
Then, for $i < G$, 
$$\text{Pe}_i = \frac{2}{T}\protect\int_0^{T/2} Pe_i(t) \text{d}t = \frac{2V_T f_b l_0}{ \pi r_0^2 D }\left( \frac1{2h}\right)^i$$ 
and for $i \geqslant G$, 
$$\text{Pe}_i = \frac{2}{T}\protect\int_0^{T/2} Pe_i(t) \text{d}t = \frac{2V_T f_b l_0}{ \pi r_0^2 D }\left( \frac1{2h}\right)^{G-1} \left( \frac{1}{2}\right)^{i-G+1} $$.

\subsection{Generation $k$ of the transition between convection and diffusion}
\label{fn13}

The generation $k$ at which the transition between convection and diffusion occurs is computed by solving the equation $\rm{Pe}_k = 1$.
If $k < G$, we have,  
$$
2^k = \left( \frac{2V_T l_0 f_b}{\pi r_0^2 D} \right)^{\frac32} = \left( 2\ve \frac{l_0}{\pi r_0^2 D} \right)^{\frac32} \propto \ve^{\frac32} \times M^{-\frac34} 
$$
and if $k\geqslant G$, 
\begin{align*}
2^k = & \ \frac{2V_T l_0 f_b}{\pi r_0^2 D} \left(2^{G-1} \right)^{\frac{1}{3}}\\ 
= & \ 2\ve \frac{l_0}{\pi r_0^2 D} \left(2^{G-1} \right)^{\frac{1}{3}} \propto \ve \times M^{-\frac{5}{24}} 
\end{align*}

\subsection{Localization of the transition from convective to diffusive transport at BMR and MMR}
\label{fn14}

At BMR, the generation index $k_{\rm{BMR}}$ at which the transition between a transport by convection and a transport by diffusion is localized depends on metabolic rate and on mammal mass,
$$
k_{\rm{BMR}} = 
\left|
\begin{array}{ll}
G - 1 + 3.41 - 0.47 \ \frac{\log(M)}{\log(2)}& (M \geqslant 154 \, \text{kg})\\
G - 1 + 2.27 - 0.31 \ \frac{\log(M)}{\log(2)}& (M \leqslant 154 \, \text{kg})
\end{array}
\right.
$$
The transition occurs in the convective tree for mammals with a mass larger than $154$ kg and in acini for mammals with a mass lower than $154$ kg.
In each compartment, the index depends linearly on the logarithm of the mass of the animal.

At MMR, the transition always occurs in acini and the corresponding generation index $k_{\rm{MMR}}$ depends linearly on the logarithm of the mass of the animal,
$$
k_{\rm{MMR}} =  G - 1 + 5.02 - 0.24 \ \frac{\log(M)}{\log(2)}
$$

The dependence of the indices $k_{\rm{BMR}}$ and $k_{\rm{MMR}}$ on mammal masses are plotted in Figure \ref{fig:convDiff}.

\section{Model equations}
\label{I}

The transport of oxygen and carbon dioxide in the lung is driven by three main phenomena: convection, diffusion and exchange with the acini walls. 
The airways are modelled as cylinders. 
We assume that the airways and fluid properties are the same in all the branches with the same generation index, hence we can study only one airway in each generation.
For the generation $i$, we define $C_i(t,x)$ as the mean oxygen concentration at the time $t$ over the slice of the cylinder located at the position $x$ on the axis of the cylinder.
Equivalently, we define the mean partial pressure $P_i(t,x)$, which is proportional to the mean oxygen concentration.

\begin{figure}[h!]
 \begin{center}
\begin{tikzpicture}[scale=0.6,rotate=90]
	\pgfmathsetmacro{\persp}{0.2};
	\pgfmathsetmacro{\Ri}{1.5}
	\pgfmathsetmacro{\Re}{2}
	\pgfmathsetmacro{\H}{2} 
	\pgfmathsetmacro{\e}{0.5}
	\draw (0,\H) ellipse ({\Re} and \Re*\persp);
	\draw (\Re,0) arc (0:-180:{\Re} and \Re*\persp);
	\draw (-\Re,0)--++(0,\H);
	\draw (\Re,0)--++(0,\H);
	\draw[dashed] (\Re,0) arc (0:180:{\Re} and \Re*\persp);
	\draw [>=latex,<->] (2.25,0)--++(0,\H);
	\draw (2.6,1.4) node[right]{$dx$}; 
	
	\draw[>=latex,->,line width=3pt] (0,3.5)--(0,2.5);
	\draw (0.6,4) node[right,scale=1.5] {$Q_l$}; 
	
	\draw[>=latex,->,line width=3pt] (0,-0.5)--(0,-1.5);
	\draw (0.6,-0.5) node[right,scale=1.5] {$Q_r$};
	
	\draw[>=latex,->,line width=3pt] (-2.25,1)--(-3.25,1);
	\draw (-2.65,0.9) node[right,scale=1.5] {$Q_w$};
	
\end{tikzpicture}
\caption{Mass balance in a slice of an idealized airway (cylinder). The variation of oxygen concentration in the slice depends on the balance between the oxygen flow entering the slice and getting out of the slice.}
\label{fig:cylindre}
 \end{center}
 \end{figure}
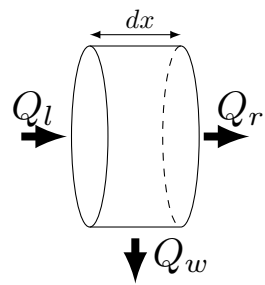

The equations of oxygen transport in a cylinder are derived using a mass balance for oxygen in a slice with thickness $dx$ localized at the position $x$ on the cylinder axis, as schematized in Figure~\ref{fig:cylindre}.

The quantity of oxygen entering the slice by the "left" side $Q_l$ in Figure~\ref{fig:cylindre} is 
$$ 
Q_l(t,x) = \left( u_i(t)C_i(t,x) - D \frac{dC_i}{dx}(t,x) \right) \pi r_i^2
$$ 
where $D$ is the diffusion coefficient of oxygen in air, $u_i(t)$ is the mean velocity of the fluid in the generation $i$ and $r_i$ is the radius of the airways of generation $i$.
The quantity of oxygen leaving the slice by the "right" side $Q_r$ in Figure~\ref{fig:cylindre} is
$$ 
Q_r(t,x) = -\left( u_i(t)C_i(t,x+dx) - D \frac{dC_i}{dx}(t,x+dx) \right) \pi r_i^2 
$$

Finally, the quantity of oxygen exchanged with the bronchus walls is
$$ 
Q_w(t,x) = - \alpha_i \rho_s \left( P_i(x) - \Pb \right) 2\pi r_i dx 
$$ 
where $\Pb$ is the O$_2$ partial pressure in blood, $\rho_s$ is the amount of exchange surface area per unit of alveolar duct surface area, see Appendix \ref{fn8}, and $\alpha_i$ is the permeability of the alveolar membrane \parencite{felici_physics_2003}:
\begin{equation}\label{eq:alpha}
  \alpha_i = \left\{
      \begin{aligned}
        & 0 & \quad  (i =  0 ... G-1) \\
        & \alpha = \frac{D_{\text{O$_2$,H$_2$O}} \sigma_{\text{O$_2$,H$_2$O}}}{\tau} & \quad (i = G...N-1)
      \end{aligned}
    \right.
\end{equation}
where  $D_{\text{O$_2$,H$_2$O}}$ is the diffusion coefficient of oxygen in water, $\sigma_{\text{O$_2$,H$_2$O}}$ is the solubility coefficient of oxygen in water and $\tau$ is the thickness of the alveolar membrane. 

\bigskip

Finally, the variation in the slice of oxygen concentration over time is 
$$ 
\pi r_i^2 dx \frac{\partial C_i}{\partial t}(t,x) = Q_l(t,x) + Q_r(t,x) + Q_w(t,x)
$$

Making the length of the slice $dx$ go to zero, we obtain for $x \in [0,l_i]$, 
\begin{equation*}
\frac{\partial C_i}{\partial t} \pi r_i^2 - \underbrace{D \frac{\partial^2 C_i}{\partial x^2}\pi r_i^2}_{\text{diffusion}} + \underbrace{u_i(t)\frac{\partial C_i}{\partial x} \pi r_i^2}_\text{convection} 
+ \underbrace{\alpha_i \rho_s \left(P_{i} - \Pb \right) 2\pi r_i}_{\text{exchange with blood}} = 0.
\end{equation*}

As concentration and partial pressure are proportional, we can work with partial pressure only. 
Finally, the transport dynamics of oxygen partial pressure in a single branch is, for $x \in [0,l_i]$,

\begin{equation}\label{EqL}
\frac{\partial P_i}{\partial t} - \underbrace{D \frac{\partial^2 P_i}{\partial x^2}}_{\text{diffusion}} + \underbrace{u_i(t)\frac{\partial P_i}{\partial x}}_{\text{convection}} + \underbrace{\beta_i \left(P_{i} - \Pb \right)}_{\text{exchange with blood}} = 0
\end{equation}

The exchange coefficient $\beta_i$ is 
\begin{equation}\label{eq:betai}
 \beta_i =  \left\{
      \begin{aligned}
        & 0 & \quad  (i = 0...G-1) \\
        & \rho_s \frac{2 k}{r_A} \alpha_i & \quad  (i = G...N-1)
      \end{aligned}
    \right.
\end{equation}
where $k$ is the ratio relating oxygen partial pressure to its concentration in water and $r_A$ is the radius of the branches in the acinus. 

\section{Blood partial pressures}
\label{II}

Blood partial pressure $P_{\rm{blood}}$ of oxygen depends non linearly on the local value of $P_i$, as a result of a balance between the amount of oxygen exchanged through the alveolar--capillary membrane and the amount of oxygen stored or freed during the passage of blood in capillaries~\parencite{noel_interplay_2019}.

As oxygen is stored within haemoglobin and dissolved in plasma, this balance writes
\begin{equation}
\alpha \left(P_{i} - P_{\rm{blood}} \right) = 4 Z_0 (f(P_{\rm{blood}})-f(\tilde{P}_{aO_2})) \\
+ \sigma_{O_2} v_s \left( P_{\rm{blood}} - \tilde{P}_{aO_2}\right)
\end{equation}
with $Z_0$ the haemoglobin concentration.
Each of haemoglobin molecules contains four sites of binding with oxygen molecules, hence the $4$ in factor of $Z_0$. 
The function $f(x) = x^{2.6}/(x^{2.6}+26^{2.6})$ is the Hill's equation~\parencite{hill_oxygen_1936} that reproduces the saturation of haemoglobin depending on oxygen partial pressure in blood. 
The quantity $v_s$ corresponds to blood velocity in capillaries and $\sigma_{O_2}$ corresponds to the solubility coefficient of oxygen in blood. 
The pressure $\tilde{P}_{aO_2} = 88$ mmHg is the effective partial pressure of oxygen in arterial lung circulation (low oxygenated blood) that accounts for potential previous visits of other alveoli by blood, as defined in~\parencite{noel_interplay_2019}. 
This quantity is assumed independent of the mammal species~\parencite{lindstedt_pulmonary_1984}.

The mean blood velocity $v_s$ depends on the mass and on the metabolic regime studied. 
It can be computed as the ratio of the capillary length $l_c$ over the transit time in a capillary $t_c$.
As in \textcite{west_general_1997}, we assume that the terminal units of the blood network are invariant in size. 
Hence, the capillary length is constant in our model and equals to $1 \, \text{mm}$. 
The transit time in capillaries depends both on mass and on metabolic rate,
$$
\begin{array}{ll}
t_c \simeq 0.36 \ M^{\frac14} & \text{at basal metabolic rate~\parencite{west_general_1997, haverkamp_physiologic_2005}}\\
t_c \simeq 0.25 \ M^{0.165} & \text{at maximal metabolic rate~\parencite{bishop_integration_2013, haverkamp_physiologic_2005}}
\end{array}
$$
No data is available in the literature for field metabolic rate. 
Nevertheless, we determine a default allometric scaling law based on the fact that field metabolic rate~\parencite{hudson_relationship_2013} is more similar to basal metabolic rate than maximal metabolic rate for which the energy is mostly spent by muscle activity~\parencite{haverkamp_physiologic_2005}.
Hence, we assume that the exponent for $t_c$ is the same at field metabolic rate and at basal metabolic rate.

Then, using the estimated value $t_c = 838$ s for human based on the data from Haverkamp et al. \parencite{haverkamp_physiologic_2005}, we use as the allometric scaling law for the transit time at field metabolic rate, 
$$t_c \simeq 0.29 \ M^{\frac14} \, \, \text{at field metabolic rate }$$
Notice that the model sensitivity relatively to this hypothesis is very low, as indicated in Appendix \ref{VI}. 

\section{Boundary conditions}
\label{III}

To mimic lung bifurcations, we use continuity conditions $P_i(l_i,t) = P_{i+1}(0,t)$ and conservation of the number oxygen molecules
\begin{equation}
 S_i\left( u_i(t)P_i(t,l_i) - D\frac{\partial P_i(t,l_i)}{\partial x}\right)
- 2 S_{i+1} \left(u_{i+1}(t)P_{i+1}(t,0) - D\frac{\partial P_{i+1}(t,0)}{\partial x} \right) = 0
\end{equation}
The $2$ on the righthandside of the last expression indicates that an airway in the generation $i$ bifurcates into two airways in the generation $i+1$.

Oxygen conservation can be rewritten, using the previous continuity condition, 
\begin{equation}
 - DS_i\frac{\partial P_i(t,l_i)}{\partial x} = - 2 D S_{i+1} \frac{\partial P_{i+1}(t,0)}{\partial x}
\end{equation}

Finally, we assume that $P_0(t,0) = P_{\text{air}}$ at the trachea entrance, where $P_{\text{air}}$ is the partial pressure of oxygen in ambient air. 
The surface area represented by the outlets of the deepest airways of the tree is negligible relatively to the whole exchange surface area.
Moreover, the exchange occurring at these outlets is negligible relatively to the exchange occurring in the upper parts of the acini.
Hence, we can assume that no exchange occurs at the outlets of the deepest airways in acini, i.e.
$
 - D \frac{\partial P_{N-1}}{\partial x}(l_A,t)   = 0
$.

\section{Initial conditions}
\label{IV}

At time $t=0$, we assume $u_i(0) = 0$, $\frac{\partial P_i}{\partial t }(0,x) = 0$ in the convective part of the tree ($i = 0...G-1$) and $P_i(0,x)$ constant in acini ($i = G...N-1$). 
Then, an explicit stationary solution in the bronchial tree can be derived and used as a non trivial initial condition, for $i=0...G-1$ with $P_{\text{blood}}$ fixed to $\tilde{P}_{aO2} = 88$ mmHg (see Appendix \ref{II}),
\begin{equation*}
P_i(0,x)=P_{\text{air}} + \frac{P_{\text{blood}}-P_{\text{air}}}{\sum\limits_{k=0}^N \left(\frac{1}{2h}\right)^k} \left(\sum\limits_{k=0}^{i-1} \left(\frac{1}{2h}\right)^k + \left(\frac{1}{2h} \right)^i \frac{x}{l_i} \right).
\end{equation*}
For $i=G...N-1$, we suppose that the partial pressure is the same as in blood, $P_i(0,x) = P_{\text{blood}}$.

This initial condition allows to speed up the algorithm by giving a non-trivial and physically relevant oxygen distribution at the start of the algorithm.
Nevertheless, it is necessary to run the model of oxygen transport for several ventilation cycles to reach periodic oxygen profiles in airways. 

\section{Numerical scheme}
\label{V}

This model is analyzed with numerical simulations that allow to get numerical approximation of the solutions of the equations system. 
The numerical method is based on a discretization of the transport equations using an implicit finite differences scheme. 
The computation are performed using the computing language Julia \parencite{bezanson_julia_2017}. 
From the initial distribution of partial pressures in the tree, the simulations are then run up to a time when the oxygen concentration pattern becomes periodic in time.
 All the model predictions are based on computations made when the oxygen profile is periodic. 

The optimization process is made by inverting numerically the implicit constraint $f_{O_2}(V_T, f_b) = \vo$ with the secant method. 
The inversion is equivocal and allows to compute numerically the non-linear function $f_b \rightarrow V_T(f_b)$. 
Then, the optimization is performed on the unidimensional function $f_b \rightarrow  \tilde{\mathcal{P}}(V_T(f_b),f_b)$ by computing explicitly $\frac{d \tilde{\mathcal{P}}(V_T(f_b),f_b)}{d f_b}$ from equation (\ref{power}) and by solving $\frac{d \tilde{\mathcal{P}}(V_T(f_b),f_b)}{d f_b} = 0$.
The derivative $\frac{d \tilde{\mathcal{P}}(V_T(f_b),f_b)}{d f_b}$ depends on $V_T(f_b)$ and $\frac{d V_T(f_b)}{d f_b}$.
The quantity $\frac{d V_T(f_b)}{d f_b}$ is estimated numerically using the approximation $\frac{d V_T(f_b)}{d f_b} = \frac{V_T(f_b + m) - V_T(f_b)}{m}$ with $m$ a scalar small relatively to $f_b$. 

\section{Sensitivity analysis}
\label{VI}

Running sets of simulations, we studied the parameters sensitivity of our model, more specifically for the parameters for which the data in the literature are scarce or missing.

First, our sensitivity analysis shows that our model has a very low sensitivity to the allometric scaling law of the blood residence time in pulmonary capillaries, indicating that the choice made for the transit time of blood at field metabolic rate does not affect significantly the model predictions.

The hydrodynamic resistance $R$ is positively correlated to the exponent of breathing rate $f_b$.  
A hydrodynamic resistance independent of the ventilation regime leads to good predictions for breathing rates at both BMR and MMR. 
This hypothesis is supported by the reported changes in dead volume during exercise and by the effects of inertia and turbulence on the hydrodynamic resistance~\parencite{haverkamp_physiologic_2005}.
Indeed, if we neglect the inertia and turbulence in the bronchi at MMR, the change in dead volume at this regime leads the hydrodynamic resistance to be decreased by a factor larger than $3$. 
In this case, the corresponding exponent for breathing rates drops to $-0.10$. 
Consequently inertia and turbulence might play an important role on the control of breathing rates, but, interestingly, their influence seems to be balanced by the dead volumes increase.
Hence, this shows that the hypothesis of a hydrodynamic resistance independent of the ventilation regime is a satisfactory approximation.

\end{document}